\documentclass[12pt]{article}

\usepackage{amssymb,amsfonts,amsmath,amsthm}
\usepackage{mathtools}
\usepackage{setspace}
\usepackage{sourceserifpro}
\usepackage[T1]{fontenc}
\usepackage{graphicx}
\usepackage[left=1in,right=1in,top=1.5in,bottom=1.5in]{geometry}
\usepackage[authoryear]{natbib}
\usepackage{enumerate}
\usepackage{hyperref}
\usepackage{booktabs}
\usepackage{pdflscape}
\usepackage{tabularx}
\usepackage{longtable}
\usepackage{epstopdf}
\usepackage[utf8]{inputenc}
\usepackage{multirow}
\usepackage{subcaption}
\usepackage[flushleft]{threeparttable}
\usepackage{enumitem}
\usepackage{comment}
\usepackage{soul,xcolor}
\setstcolor{red}
\usepackage{chngcntr} 
\usepackage{appendix}

\usepackage{geometry}
\usepackage{tikz}
\usetikzlibrary{positioning}

\usepackage{bbm}
\usepackage{forest}
\usepackage{float}
\usepackage{rotating}

\usepackage{algorithm}
\usepackage{algpseudocode}

\hypersetup{
colorlinks=true,
linkcolor=blue,
filecolor=magenta,
urlcolor=blue,
citecolor=blue,
}
\def\sym#1{\ifmmode^{#1}\else\(^{#1}\)\fi}

\newtheorem{theorem}{Theorem}[section]
\newtheorem{lemma}[theorem]{Lemma}
\newtheorem{proposition}[theorem]{Proposition}
\newtheorem{definition}[theorem]{Definition}

\newcommand{\appendixpagenumbering}{
	\break
	\pagenumbering{arabic}
	\renewcommand{\thepage}{\thesection-\arabic{page}}
}

\begin{document}
\setstretch{1.5}
\begin{titlepage}
\title{Behavioral Machine Learning? Regularization and Forecast Bias}
\author{Murray Z. Frank, Jing Gao, and Keer Yang\thanks{Frank is from University of Minnesota, Minneapolis, MN, E-mail: murra280@umn.edu. Gao is from University of Minnesota, Minneapolis, MN, E-mail: gao00268@umn.edu. Yang is from University of California at Davis, Davis, CA, E-mail: kkeyang@ucdavis.edu. None of the authors have any relevant or material financial interests related to the research described in this paper. We are grateful for helpful comments from Jiangze Bian, Nirmol Das, Melina Ludolph (discussant), Ke Tang (discussant), Rui Da (discussant), Colin Ward, Andy Winton, Moto Yogo and seminar participants at the University of Minnesota, 2023 JFDS Conference, 2023 ABFER, HKUST, and SUSTech Research Conference on Capital Market Research in the Era of AI, 2023 Cardiff Fintech Conference, 2023 FMA, 2023 KAIST AI Social Science BootCamp, 2024 UC Davis Finance Day, 2024 Econometric Society conference on Economics and AI+ML, 2025 CICF, and 2025 Financial Intermediation in the 3rd Millennium conference at Essex.  An earlier draft of this paper circulated with the title, ``Behavioral Machine Learning? Computer Predictions of Corporate Earnings also Overreact''. We alone are responsible for all errors.}}

\date{November 28, 2025}
\maketitle
\thispagestyle{empty}

\begin{abstract}
\begin{singlespace}
Standard forecast efficiency tests interpret violations as evidence of behavioral bias. We show theoretically and empirically that rational forecasters using optimal regularization systematically violate these tests. Machine learning forecasts show near zero bias at one year horizon, but strong overreaction at two years, consistent with predictions from a model of regularization and measurement noise. We provide three complementary tests: experimental variation in regularization parameters, cross-sectional heterogeneity in firm signal quality, and quasi-experimental evidence from ML adoption around 2013. Technically trained analysts shift sharply toward overreaction post-2013. Our findings suggest reported violations may reflect statistical sophistication rather than cognitive failure.
\end{singlespace}
\end{abstract}

JEL Classification: G17, G32, G40

Key Words: Overreaction, underreaction, regularization, machine learning, behavioral finance 





\clearpage
\end{titlepage}

\DeclareGraphicsExtensions{.pdf,.png,.gif,.jpg}
\newpage
\section{Introduction}

Since the introduction of rational expectations \citep{Muth1961,lucas1972expectations}, there have been extensive tests of whether actual forecasts satisfy efficiency conditions predicted by rational expectations. Professional forecasts of corporate earnings, inflation, and GDP growth routinely violate the prediction that forecast errors should be unpredictable using information available when forecasts were made. The \citet{coibion2015information} test has become the main approach to testing. It regresses forecast errors on forecast revisions. Forecasts are said to be efficient if the forecast errors are unpredictable using information available when forecasts were made. In many contexts, the application of this test results in strongly rejecting the null hypothesis of forecast efficiency. Both underreaction (forecasts are revised too cautiously), and overreaction (the revisions overshoot) have been widely reported in a range of settings \citep{coibion2015information,bordalo2022overreaction,afrouzi2023overreaction,de2024noise}.

The dominant interpretation in the literature attributes these violations to behavioral biases \citep{hirshleifer2015behavioral,barberis2018psychology,bordalo2024belief}. The documented forecast biases are said to be due to human cognitive limitations, brain structure, and emotions. Forecasters may anchor on prior beliefs \citep{kahneman2013prospect}, extrapolate excessively \citep{DeBondt1985}, overweight salient signals \citep{bordalo2020overreaction},  underreact due to cognitive constraints \citep{sims2003implications}, or have noisy expectations \citep{de2025expectations}. These behavioral views have become influential in both finance and macroeconomics. They are important to our understanding of expectation formation and are increasingly being embedded into models of asset prices and business cycles. The interpretation seems convincing. If rational forecasters use all available information optimally, forecast errors should be unpredictable. Systematic violations must therefore reflect departures from rationality.

We applied the standard tests \citep{coibion2015information} to predictions generated by various popular machine learning (ML) algorithms \citep{hastie2009elements}. Like the predictions generated by people, the machine learning generated predictions also reject the efficiency null hypothesis. But of course, these ML algorithms do not have human brain structure nor do they have human emotions.  

We therefore turn to an alternative interpretation. We show that optimal statistical procedures produce forecasts that systematically violate the efficiency test null hypothesis. Forecasters facing noisy signals rationally regularize their estimates. They sensibly shrink coefficients toward zero in order to reduce variance at the cost of introducing bias. This bias-variance tradeoff has been understood since \citet{Stein1956}, which showed that biased shrinkage estimators dominate unbiased estimators in finite samples. Modern machine learning algorithms explicitly implement this concept in widely used methods including ridge regression, random forests, and gradient boosting \citep{zou2005regularization,hastie2009elements}. We show that when forecasters optimally regularize, they systematically violate the \citet{coibion2015information} test (CG) despite behaving rationally. The CG test is a test of whether the forecasts are unbiased. The basic assumption is that unbiased estimation is optimal. When signals are sufficiently noisy, we show that this assumption fails. Consistent with \citet{Stein1956}, unbiased estimates are not optimal in that case.

Our theoretical contribution shows how optimal regularization affects forecast efficiency tests. We provide a model in which the forecasters are trying to predict an AR(1) fundamental. But they only observe a signal that is the fundamental contaminated by AR(1) noise. So they choose regularization intensity to minimize mean squared forecast error. The model produces several testable predictions that distinguish our mechanism from behavioral alternatives. Stronger regularization produces an increased CG coefficient. This shifts forecasts from overreaction toward underreaction. Lower noise persistence and higher noise volatility decrease the coefficient, amplifying overreaction. At longer forecast horizons, the effect of measurement noise is stronger. This generates negative CG coefficients even when short horizon coefficients are positive. Behavioral theories would require horizon dependent bias parameters to match this pattern. Our mechanism produces these patterns naturally from the interaction of measurement noise and optimal shrinkage.

Our empirical strategy uses analyst earnings forecasts and machine learning predictions to test our regularization interpretation. We focus primarily on three ML methods: ridge regression, gradient boosting, and random forests. We use standard historical data to predict corporate earnings. These algorithms implement explicit regularization. As mentioned, the ML forecasts exhibit systematic CG test violations that are certainly not identical, but they are often similar to human analysts. The horizon pattern is striking and the difference between humans and the machines is interesting. At one year horizons, ML forecasts show near zero coefficients while human analysts exhibit some underreaction. At two year horizons, ML forecasts show strong overreaction, compared to near zero for humans. This systematic sign change precisely matches our theoretical predictions and is not predicted by uniform psychological biases.

We use three complementary identification strategies to distinguish regularization from behavioral explanations and to account for the horizon effects. First, we experimentally vary regularization intensity within ML models. The CG coefficient responds monotonically as theory predicts. This controlled variation provides direct evidence that regularization drives violations with reasonable magnitudes. 

Second, we exploit cross-sectional variation in signal quality. Some firms have noisier signals than do others. We measure this using high R\&D intensity and young age. Such firms are associated with  more negative CG coefficients under our theory but not under uniform behavioral bias. In the data we find that high R\&D firms have coefficients that are more negative than low R\&D firms. Young firms have CG coefficients that are more negative than old firms. These patterns hold for both human and ML forecasts. This supports the idea that the interaction of measurement noise and regularization rather than psychology drives the heterogeneity. 

Third, we use a quasi-natural experiment exploiting the variation from ML adoption timing. The Python scikit-learn library reached its first stable version in 2012. That made regularized forecasting accessible even without a heavy programming investment of time and effort. We manually collect analyst educational backgrounds from LinkedIn and FINRA records, classifying 173 analysts as technical (statistics, mathematics, computer science degrees) and 685 as non-technical. Pre-2013, technical analysts show near zero CG coefficients, while non-technical analysts show significant underreaction. Post-2013, technical analysts shift sharply to overreaction, while non-technical analysts show only a modest change. This difference in responses by technical training provides extra evidence that statistical methods rather than psychology, drives forecast patterns. 

The economic magnitudes that we document are substantial. ML methods achieve 7\% lower mean squared forecast error than human analysts over the full period 1986 to 2019. Regularization does improve accuracy. The performance gap reverses post-2013. This is consistent with human analysts increasingly adopting ML methods. Despite their competitive accuracy, ML forecasts violate efficiency tests as strongly as do humans. The CG test violations and forecast quality are distinct dimensions. Following the approach of \citet{bordalo2025real} we show these violations affect real decisions. The predicted forecast errors from technical analysts strongly predict subsequent investment reversals. This suggests managers respond to analyst forecasts when allocating capital, amplifying any systematic patterns from regularization.

Our results contribute to several literatures. Most directly, we connect to the debate on forecast efficiency and expectation formation. A large literature documents violations of rational expectations tests and attributes them to behavioral biases. \citet{DeBondt1985} and \citet{Lakonishok1994} document overreaction in analyst forecasts. They attribute this to representativeness heuristics. \citet{barberis1998model} develops a model in which investors underweight information due to conservatism bias. \citet{bordalo2020overreaction} propose diagnostic expectations, in which agents overweight salient signals. Our contribution shows violations can arise from optimal statistical practice, requiring more refined tests to distinguish mechanisms. \citet{de2024noise} provide important recent evidence that analyst forecast errors contain substantial transient noise, generating negative CG coefficients. We complement this by modeling how rational forecasters should optimally respond to noise through regularization. We also provide testable comparative statics. Our simulation experiments, cross-sectional, and quasi-experimental evidence goes beyond just showing that noise exists. We characterize the optimal response and tests whether forecasters implement it.

Several papers have proposed non-behavioral explanations for forecast efficiency violations due to information frictions \citep{sims2003implications,mankiw2002sticky,woodford2020modeling}. The information friction models typically predict underreaction. They do not naturally explain why machine learning algorithms exhibit similar violations to humans. Nor do they explain why violations flip sign across horizons. \citet{martin2022market} study how regularization affects market efficiency tests in high dimensional settings. They find that penalized regression can alter inference about return predictability. Our focus differs. We study forecast efficiency tests rather than market efficiency. We provide direct evidence through ML adoption timing and analyst heterogeneity. 

A growing literature uses machine learning for prediction in finance \citep{gu2020empirical,kelly2019characteristics,chinco2019sparse}. \citet{zhang2025man} compare ML forecasts to human analysts for earnings. They find competitive accuracy. We build on their methodology but focus on whether ML forecasts violate rational expectations tests, and what this implies about the source of violations. Our contribution differs by examining equilibrium consequences when forecasters adopt ML methods. We show that reported violations may reflect statistical sophistication rather than cognitive failure.

Overall, our results have implications for interpreting forecast efficiency tests and understanding expectation formation. Violations do not necessarily indicate cognitive failure nor behavioral bias. They may reflect optimal statistical methods applied to noisy signals. Distinguishing these explanations requires tests exploiting variation in regularization intensity and signal quality. As machine learning gains adoption in financial markets, forecast patterns should shift predictably. Explicit regularization reduces short horizon errors but amplifies long horizon mean reversion. Evaluating forecast quality requires examining out of sample performance rather than in sample efficiency test satisfaction. If test interpretations are based on incorrect assumptions about the data generating process, inferences about rationality can be misleading.

The rest of the paper is organized as follows. Section \ref{sec:theory} presents our theory showing how regularization affects forecasts. Section \ref{sec:data} provides the data sources and describes the basic empirical methodology. Section \ref{sec:evidence} gives our evidence. The conclusion is in Section \ref{sec:conclusion}. 


\section{Theoretical Framework}
\label{sec:theory}

In this section we develop a model of optimal forecasting under regularization to characterize violations of the \citet{coibion2015information} (CG) forecast efficiency test. Our analysis shows that rational statistical procedures systematically generate patterns that appear as behavioral biases in standard tests.

\subsection{Environment}

Consider a forecaster predicting outcomes $y_{t+1}$ using noisy signals $z_t$. The data generating process is as follows.

\textbf{Outcomes.} Realized outcomes depend on an unobserved fundamental $s_t$ plus idiosyncratic noise.
\begin{equation}
y_{t+1} = \alpha s_{t+1} + \epsilon_{t+1},
\end{equation}
where $\alpha > 0$ is a known scaling parameter and $\epsilon_{t+1} \sim \text{i.i.d.}(0, \sigma_\epsilon^2)$ is independent of all other variables.

\textbf{Fundamentals.} The fundamental follows a stationary AR(1) process.
\begin{equation}
s_{t+1} = \rho_s s_{t} + v_{t+1}, \quad v_{t+1} \sim \text{i.i.d.}(0, \sigma_v^2), \quad |\rho_s| < 1,
\end{equation}
with unconditional variance $\sigma_s^2 = \sigma_v^2/(1-\rho_s^2)$.

\textbf{Signals.} Forecasters observe signals contaminated by measurement noise.
\begin{equation}
z_t = s_t + \eta_t,
\end{equation}
where noise also follows an AR(1):
\begin{equation}
\eta_t = \rho_\eta \eta_{t-1} + u_t, \quad u_t \sim \text{i.i.d.}(0, \sigma_u^2), \quad |\rho_\eta| < 1,
\end{equation}
with variance $\sigma_\eta^2 = \sigma_u^2/(1-\rho_\eta^2)$. All innovations $(v_t, u_t, \epsilon_t)$ are mutually independent, implying $\text{Cov}(s_t, \eta_s) = 0$ for all $t,s$.

Empirically, noise terms are typically not persistent and often resemble white noise, implying $\rho_\eta = 0$. In this setting, we retain the AR(1) structure but assume that $\rho_\eta$ is close to zero, meaning that $\rho_s \gg \rho_\eta$. This setup is about earnings forecasting where fundamentals (true earning power) are persistent but imperfectly observed. The signal-to-noise ratio $\sigma_s^2/\sigma_\eta^2$ measures information quality. Persistence parameters $\rho_s$ and $\rho_\eta$ determine whether shocks have lasting effects. 

\subsection{Optimal Forecasting with Regularization}
At time $t$, forecasts observe $z_{t}$ and use it to forecast $y_{t+1}$. The forecast is given by
\begin{equation}
    F_{t} y_{t+1} = \beta z_{t}
\end{equation}

\textbf{The bias-variance tradeoff.} Without regularization, the optimal one-period-ahead forecast coefficient is
\begin{equation}
\beta_{OLS} = \alpha \rho_s \frac{\sigma_s^2}{\sigma_s^2 + \sigma_\eta^2}.
\end{equation}

This is the standard Kalman filter formula, weighting the signal by its informativeness. However, when signal quality is limited or the forecaster faces estimation uncertainty, ridge regularization improves out-of-sample performance by shrinking the coefficient toward zero.
\begin{equation}
\beta_\lambda = \alpha \rho_s \frac{\sigma_s^2}{\sigma_s^2 + \sigma_\eta^2 + \lambda},
\label{eq:beta_regularized}
\end{equation}
where $\lambda \geq 0$ is the penalty parameter. 

\textbf{Optimal Forecasting.}
The optimal forecast of $y_{t+1}$ at time $t$ is therefore given by $F_{t}y_{t+1} = \beta_\lambda z_{t}$. More generally, the optimal h-period-ahead forecast is $F_{t}y_{t+h} = \rho_s^{h-1} \beta_\lambda z_{t}$. Note $\beta_\lambda < \beta_{OLS}$ whenever $\lambda > 0$. In other words, the coefficient is subject to shrinkage due to regularization.

\textbf{Why regularize?} Ridge regression optimally trades bias for variance reduction. The mean squared forecast error can be decomposed.
\begin{equation}
\text{MSE} = \text{Bias}^2 + \text{Variance}.
\end{equation}
Regularization increases bias but reduces variance. When signals are noisy or samples are small, the variance reduction dominates, yielding lower MSE. This is the textbook prescription for forecasting in high-noise environments \citep{hastie2009elements}.

\subsection{The Coibion-Gorodnichenko Test}

We have now described how forecasts are generated under the simple yet general setup above. The econometrician wants to know if these predictions are rational. To do so, we apply the CG test from \citet{coibion2015information}. 

The CG test regresses forecast errors on forecast revisions.
\begin{equation}
e_{t+1} = \gamma_0 + \gamma_{CG} \, r_t + \omega_t,
\label{eq:CG_test}
\end{equation}
where $e_{t+1} = y_{t+1} - F_{t} y_{t+1}$ is the forecast error and $r_t = F_{t} y_{t+1} - F_{t-1} y_{t+1}$ is the forecast revision. Under rational expectations with no regularization and perfect information, $\gamma_{CG} = 0$. Forecast revisions should be uncorrelated with subsequent errors. A positive coefficient is commonly described as `underreaction'. Forecasters do not revise enough when new information arrives. A negative coefficient is described as `overreaction'. 

Our contribution is to characterize the population CG coefficient when forecasters use optimal regularization. We show that forecasts produced by rational statistical procedures systematically violate the prediction that $\gamma_{CG} = 0$. The structure of the data generating process interacts with regularization to produce this. 

\subsection{Main Results}
We theoretically derive the regression coefficient for the CG test in Equation \ref{eq:CG_test}.

\begin{proposition}[CG Coefficient]
\label{prop:CG_optimal}
The CG coefficients for Equation \ref{eq:CG_test} is,
\begin{equation}
\gamma_{CG}  = \frac{(1-\rho_s^2) \lambda - \rho_s (\rho_s -\rho_\eta) \sigma_\eta^2}{(1-\rho_s^2)\sigma_s^2 + (1+\rho_s^2-2\rho_s\rho_\eta)\sigma_\eta^2}.
\label{eq:gamma_optimal}
\end{equation}
\end{proposition}

\begin{proof} 
See Appendix. The derivation expands the covariance $\text{Cov}(e_{t+1}, r_t)$ and variance $\text{Var}(r_t)$ using the AR(1) structure of fundamentals and noise, and applies independence assumptions to eliminate cross-terms. 
\end{proof} 

Proposition \ref{prop:CG_optimal} reflects two competing forces that determine the sign of $\gamma_{CG}$. The first force is the underweighting of fundamentals, which provides a positive contribution and results in patterns similar to `underreaction'. This force is captured by the term $(1-\rho_s^2)\lambda$, which is positive with regularization ($\lambda>0$). When fundamentals change, forecasters underweight the signal due to regularization. 

Consider a positive fundamental shock. The forecaster revises upward, but insufficiently. The outcome $y_{t+1}$ exceeds the forecast $F_{t}y_{t+1}$, creating a positive error. Thus, positive revisions correlate with positive errors, contributing positively to $\gamma_{CG}$.

The second force is the overreaction due to measurement noise. This provides a negative contribution. Recall $\rho_s\gg\rho_\eta$, so the term $- \rho_s (\rho_s -\rho_\eta) \sigma_\eta^2$ is negative. When noise fluctuates, forecasters respond as if fundamentals changed. 

Consider a positive shock to measurement noise $\eta_t$. The forecaster observes $z_{t} > z_{t-1}$ and revises upward. But this revision reflects noise, not fundamentals. The noise contributes negatively to $\gamma_{CG}$. This is because the noise appears in current period forecasts and is therefore related to both forecast revision and the forecast error yet in different directions. The importance of noise, while modeled differently, is also found in \citet{de2024noise}. 

The net effect depends on which force dominates. Regularization amplifies the positive force by increasing the term $(1-\rho_s^2) \lambda$. 

Special cases can help with the intuition. Suppose there is no regularization, i.e. $\lambda = 0$, we have
\begin{equation}
\gamma_{CG} = \frac{(1-\rho_s^2) \lambda - \rho_s (\rho_s -\rho_\eta) \sigma_\eta^2}{(1-\rho_s^2)\sigma_s^2 + (1+\rho_s^2-2\rho_s\rho_\eta)\sigma_\eta^2} = \frac{(\rho_\eta - \rho_s)\rho_s\sigma_\eta^2}{(1-\rho_s^2)\sigma_s^2 + (1+\rho_s^2-2\rho_s\rho_\eta)\sigma_\eta^2}.
\end{equation}
Even unregularized forecasters violate the CG test unless $\rho_\eta = \rho_s$. The sign depends on the relative persistence of the noise and the fundamental variable. The CG coefficient $\gamma_{CG} < 0$ if noise is less persistent than fundamentals ($\rho_\eta < \rho_s$), e.g. if noise is white noise, which is very likely.

Next suppose there exists only regularization but no noise. The CG coefficient becomes
\begin{equation}
\gamma_{CG} = \frac{(1-\rho_s^2) \lambda}{(1-\rho_s^2)\sigma_s^2 + (1+\rho_s^2-2\rho_s\rho_\eta)\sigma_\eta^2}.
\label{eq:gamma_optimal_no_noise}
\end{equation}
In this case, applying regularization ($\lambda > 0$) results in positive CG coefficients.

The critical point here is that optimal forecasting using regularization does not eliminate CG violations. It changes the magnitude and sometimes the sign of CG coefficients. The relative magnitudes of regularization and noise determine outcomes. The accounting differs because sophisticated forecasters partially filter persistent noise when forming revisions.

\subsubsection{Comparative Statics}\label{sec:comp}

The formula in Proposition \ref{prop:CG_optimal} provides testable predictions about how $\gamma_{CG}$ varies with economic fundamentals, regularization and noise. Below we discuss relevant comparative statics, which guide our empirical tests.

\begin{proposition}[Regularization Intensity]
\label{hyp:regularization}
As $\lambda$ increases, the CG coefficient increases.
\end{proposition}
Stronger regularization increases $(1-\rho_s^2)\lambda$. Empirically, we can test this by varying regularization intensity in machine learning models.

\begin{proposition}[Noise Persistence]
\label{hyp:noise_persistence}
The CG coefficient is more negative when noise persistence $\rho_\eta$ is low.
\end{proposition}
Arithmetically, a small $\rho_\eta$ decreases the numerator, $- \rho_s (\rho_s -\rho_\eta) \sigma_\eta^2$, and increases the denominator, $(1+\rho_s^2-2\rho_s\rho_\eta)\sigma_\eta^2$, resulting in a smaller CG coefficient. This is because transient noise ($\rho_\eta \approx 0$) fluctuates without persistence. Forecasters respond to these fluctuations. But the noise does not affect future outcomes. This creates negative correlations between revisions and errors.

\begin{proposition}[Noise Volatility]
\label{hyp:signal_quality}
The CG coefficient is more negative when noise volatility $\sigma_\eta^2$ is high.
\end{proposition}
Arithmetically, a higher $\sigma_\eta^2$ decreases the numerator, $-\rho_s (\rho_s - \rho_\eta)\sigma_\eta^2$, and increases the denominator, $(1 + \rho_s^2 - 2\rho_s\rho_\eta)\sigma_\eta^2$, resulting in a smaller CG coefficient. We assume that transient noise is empirically negligible, i.e., $\rho_\eta \approx 0$. Higher volatility leads forecasters to rely more heavily on current information, which induces a negative correlation between forecast revisions and forecast errors.

These hypotheses guide our empirical analysis. We test them using variation in regularization parameters in machine learning models, cross-sectional firm characteristics (noise persistence and volatility), and time-series variation in analyst behavior (machine learning adoption).

\subsection{Implications for Interpreting Forecast Efficiency Tests}

Our theoretical results have several implications for empirical work testing rational expectations against behavioral biases.

Most importantly, violations need not indicate irrationality on the part of forecasters. Standard interpretations treat $\gamma_{CG} \neq 0$ as evidence of behavioral bias. Our analysis shows that optimal statistical procedures generate systematic violations. Researchers must distinguish between rational regularization and behavioral bias. Optimal bias-variance tradeoff given noisy signals is not the same thing as systematically failing to use available information or overreacting emotionally. 

Cross-sectional predictions aid identification. If violations reflect regularization rather than behavioral bias, they should vary systematically with regularization intensity. Specifically, the CG coefficient should increase with stronger shrinkage and decrease with lower noise persistence and higher noise volatility. Our empirical analysis exploits these predictions to test whether observed violations are consistent with rational regularization.

The sign of violations is informative. Behavioral theories often predict uniform patterns (e.g., always underreaction or always overreaction). Our theory predicts the sign depends on persistence parameters and signal quality. This heterogeneity is a crucial distinguishing feature.

There are policy implications. If violations reflect rational statistical practice rather than behavioral bias, interventions designed to ``correct'' forecast inefficiency may be counterproductive. Forecasters may already be optimizing appropriately given their information constraints.

The next sections test these implications using analyst earnings forecasts, where we examine differences in regularization in machine learning models, cross-sectional variation in noise persistence and volatility, and analysts' technical sophistication using the timing of ML adoption.


\section{Data and Empirical Methods}
\label{sec:data}

Our empirical work uses analyst earnings forecasts from IBES, firm financial data from Compustat and CRSP, macroeconomic variables from the Federal Reserve Bank of Philadelphia, and manually collected analyst background information from LinkedIn and FINRA records. The sample covers 1986 to 2019, so that it ends before the COVID period.

\subsection{Data Sources and Sample Construction}

We use annual earnings per share (EPS) forecasts and realizations from IBES. Realized earnings come from the IBES Actual files rather than Compustat due to differences in accounting methods for measuring actual earnings. The IBES database provides firm level forecasts made by individual equity analysts. We use analyst forecasts of annual EPS at horizons of one, two, and three years. The forecasting horizon is defined as the number of months between the fiscal year end of the realized annual EPS to be predicted and the month in which the forecast is issued. For each firm and month, we compute the consensus forecast as the median of individual analyst forecasts. 

We impose standard sample restrictions. We exclude financial firms (SIC codes 6000-6999) and require non-missing data for the predictor variables. Our final sample includes 99,963 firm-year observations covering 1986 to 2019.

An important test is whether analyst technical training affects forecast patterns following ML adoption. To do that we manually collect analyst educational backgrounds using LinkedIn profiles and FINRA BrokerCheck records. We identify analysts who issued at least one earnings forecast in 2018. Then we search for their LinkedIn profiles. Cross-referencing with FINRA BrokerCheck improves matching accuracy. In total, we successfully link 858 analysts to their LinkedIn profiles.

We classify analysts as `technical' or not depending on their higher education background. We manually verify the curriculum for each major and identify technical majors as those requiring coursework in advanced statistics, linear algebra, and computational methods. The majors classified as technical are primarily STEM majors, though with some variation. For example, general biology degrees do not typically require advanced statistical courses and are therefore excluded. However, analysts with degrees in Computational Biology are classified as technical since their programs require advanced statistical tools. We also search LinkedIn profiles for self-reported technical skills including machine learning, artificial intelligence, and advanced statistics.

In total, we identify 173 analysts with technical education backgrounds and 685 non-technical analysts. For each firm month, we calculate separate median consensus forecasts for technical analysts ($F^{Tech}_t Y_{i,t+1}$) and non-technical analysts ($F^{Non-Tech}_t Y_{i,t+1}$). For this part, our sample covers 1994 to 2018, producing 14,901 firm-year observations for forecasts made by technical analysts and 36,155 firm-year observations for forecasts made by non-technical analysts.

\subsection{Machine Learning Forecasts}

To benchmark our theoretical predictions against machine learning implementations, we construct ML based earnings forecasts following the methodology of \citet{zhang2025man}. Our predictor variables include: (i) financial ratios from the Wharton Research Data Services (WRDS) Financial Ratio suites, (ii) the last available annual earnings per share from IBES, (iii) the consensus analyst forecast of EPS, and (iv) macroeconomic variables from the Philadelphia Fed's real-time dataset (Real Gross Domestic Product, Real Personal Consumption Expenditures, Industrial Production Index, and Civilian Unemployment Rate). For variables with a large number of missing observations, we replace missing values with the industry average.

We implement three regularized forecasting methods. These are ridge regression, gradient boosting, and random forests. Ridge regression shrinks coefficient estimates using an L2 penalty parameter $\lambda$ selected by cross-validation. In our theoretical framework Section \ref{sec:theory}, the optimal one period ahead forecast coefficient with regularization is
\begin{equation}
\beta_\lambda = \alpha\rho_s \frac{\sigma^2_s}{\sigma^2_s + \sigma^2_\eta + \lambda},
\end{equation}
where $\alpha > 0$ scales outcomes to fundamentals, $\rho_s$ is fundamental persistence, $\sigma^2_s$ is fundamental variance, $\sigma^2_\eta$ is noise variance, and $\lambda \geq 0$ is the regularization penalty. Note that $\beta_\lambda < \beta_{OLS}$ whenever $\lambda > 0$. 

Gradient boosting builds forecasts sequentially, with each tree correcting errors from previous trees. The learning rate controls regularization intensity. Random forests average predictions across many decision trees, with regularization operating through maximum tree depth limits and the number of trees. For each method, we estimate models using expanding windows to avoid look ahead bias.

\subsection{Variable Definitions}

\textbf{Forecast error.} For firm $i$ in fiscal year $t$, the one year ahead forecast error is
\begin{equation}
e_{i,t+1} = y_{i,t+1} - F_t y_{i,t+1}
\end{equation}
where $y_{i,t+1}$ denotes realized earnings per share from the IBES Actual files and $F_t y_{i,t+1}$ represents the consensus forecast made during fiscal year $t$ for year $t+1$ earnings. We measure $F_ty_{i,t+1}$ as the first forecast of $y_{i,t+1}$ released in fiscal year $t$ after the announcement of year $t$'s actual EPS, $y_{i,t}$. Two-year forecast errors are defined analogously.

\noindent\textbf{Forecast revision.} The forecast revision is
\begin{equation}
r_{i,t} = F_t y_{i,t+1} - F_{t-1} y_{i,t+1},
\end{equation}
the change in the consensus forecast for fiscal year $t+1$ earnings between fiscal years $t-1$ and $t$. We measure $F_{t-1}y_{i,t+1}$ as the first forecast of $y_{i,t+1}$ released in fiscal year $t-1$ after the announcement of year $t-1$'s actual EPS, $y_{i,t-1}$.

\noindent\textbf{Signal quality proxies.} Following the literature, we proxy for signal quality using firm age and R\&D intensity (R\&D expenditure scaled by total assets). Higher R\&D intensity and lower firm age are interpreted as proxies for noisier signals.

\subsection{Identification Strategy}

We use four complementary identification strategies to test our theoretical predictions.

First, we apply the \citet{coibion2015information} forecast efficiency test to both human analyst and machine learning forecasts. As mentioned previously, the CG test regresses forecast errors on forecast revisions $e_{t+1} = \gamma_0 + \gamma_{CG} r_t + \omega_t$, where $e_{t+1} = y_{t+1} - F_t y_{t+1}$ is the forecast error and $r_t = F_t y_{t+1} - F_{t-1} y_{t+1}$ is the revision. Under rational expectations with no regularization and perfect information, $\gamma_{CG} = 0$. A positive coefficient indicates underreaction, and a negative coefficient indicates overreaction.

Second, based on our theoretical framework, we systematically vary regularization intensity of ML models and examine its effect on $\gamma_{CG}$. Our theory predicts that stronger regularization increases the CG coefficient (Proposition \ref{hyp:regularization}), shifting forecasts from overreaction toward underreaction. For ridge regression, we systematically vary the penalty parameter $\alpha$. For gradient boosting, we vary the learning rate.\footnote{We vary the learning rate in gradient boosting models as a representative example of tree-based approaches.} This controlled variation provides direct evidence that regularization drives $\gamma_{CG}$ away from 0.

Third, we exploit cross-sectional variation in signal quality. Our theory predicts that firms with noisier signals exhibit more negative CG coefficients (Propositions \ref{hyp:noise_persistence} and \ref{hyp:signal_quality}). We measure signal quality using R\&D intensity and firm age. We interact forecast revisions with firm characteristics
\begin{equation}
e_{i,t+1} = \gamma_1 r_{i,t} + \gamma_2 r_{i,t} \times \text{Characteristic}_i + \alpha_i + \delta_t + \varepsilon_{i,t+1}.
\end{equation}
These patterns should hold for both human and ML forecasts if regularization rather than psychology drives heterogeneity.

Fourth, we use a quasi-natural experiment exploiting variation in ML adoption timing. The Python scikit-learn library reached its first stable version in 2012, making regularized forecasting widely accessible to practitioners. This library includes ridge regression, LASSO, elastic net, random forests, and gradient boosting with standardized APIs. We exploit variation in analyst technical training to identify differential ML adoption. Our prediction is that analysts with quantitative training should adopt ML methods earlier and more aggressively, exhibiting sharper changes in forecast patterns post-2013.

\subsection{Summary Statistics}

Table \ref{tab:summary} reports summary statistics for the main sample. Panel A shows firm characteristics including EPS, book-to-market ratio, return on assets, and debt-to-assets ratio. Panel B shows forecast levels for human analysts and the three ML methods. Mean forecasts range from 1.322 (random forests) to 1.436 (human analysts). Panel C shows forecast errors, defined as realized minus predicted EPS. The mean errors are -0.214 for human analysts, -0.169 for ridge regression, -0.110 for gradient boosting, and -0.100 for random forests over the full 1986 to 2019 period, suggesting that all are overly pessimistic. Panel D shows forecast revisions, defined as the EPS forecast at time $t$ minus that at time $t-1$. The mean revisions are -0.266 for human analysts, -0.053 for ridge regression, -0.020 for gradient boosting, and -0.003 for random forests over the full 1986 to 2019 period. 

\begin{table}[htbp!]
\centering
\caption{\textbf{Summary Statistics}}
\label{tab:summary}
\begin{threeparttable}
\small
\begin{tabular}{lcccccccc}
\toprule
Variable & Mean & Std Dev & P25 & Median & P75 & Min & Max & N\\
\midrule
\multicolumn{9}{l}{\textit{Panel A: Firm Characteristics}} \\
EPS           & 1.222 & 2.022 & 0.130 & 0.990 & 2.070 & -9.547 & 18.276 & 99963 \\
Book to Market            & 0.618 & 0.473 & 0.302 & 0.518 & 0.806 & 0.014  & 6.813  & 99963 \\
Return on Assets           & 0.082 & 0.198 & 0.030 & 0.110 & 0.177 & -1.491 & 0.581  & 99963 \\
Debt to Assets      & 0.214 & 0.200 & 0.042 & 0.172 & 0.334 & 0.000  & 1.139  & 99963 \\
\midrule
\multicolumn{9}{l}{\textit{Panel B: Forecast Levels}} \\
Human Analyst Forecast & 1.436 & 1.784 & 0.400 & 1.150 & 2.150 & -6.417 & 18.331 & 99963 \\
Ridge Forecast & 1.391 & 1.840 & 0.321 & 1.120 & 2.147 & -7.389 & 18.932 & 99963 \\
Gradient Boosting Forecast & 1.332 & 1.758 & 0.286 & 1.078 & 2.073 & -7.360 & 15.776 & 99963 \\
Random Forest Forecast & 1.322 & 1.805 & 0.251 & 1.070 & 2.088 & -7.047 & 15.992 & 99963 \\
\midrule
\multicolumn{9}{l}{\textit{Panel C: Forecast Errors}} \\
Human Forecast Error       & -0.214 & 0.956 & -0.369 & -0.050 & 0.120 & -21.951 & 11.160 & 99963 \\
Ridge Forecast Error       & -0.169 & 0.943 & -0.356 & -0.045 & 0.168 & -22.165 & 11.375 & 99963 \\
Gradient Boosting Error    & -0.110 & 0.943 & -0.287 & 0.001  & 0.206 & -19.651 & 11.235 & 99963 \\
Random Forest Error        & -0.100 & 0.939 & -0.270 & 0.011  & 0.209 & -19.856 & 11.212 & 99963 \\
\midrule
\multicolumn{9}{l}{\textit{Panel D: Forecast Revisions}} \\
Human Forecast Revision    & -0.266 & 0.784 & -0.450 & -0.100 & 0.050 & -13.334 & 9.571  & 99963 \\
Ridge Forecast Revision    & -0.053 & 1.455 & -0.438 & -0.008 & 0.356 & -19.856 & 45.156 & 99963 \\
Gradient Boosting Revision & -0.020 & 0.901 & -0.323 & 0.037  & 0.341 & -12.974 & 10.374 & 99963 \\
Random Forest Revision     & -0.003 & 0.917 & -0.289 & 0.058  & 0.348 & -13.360 & 10.023 & 99963 \\
\bottomrule
\end{tabular}
\begin{tablenotes}
\small
\item This table reports summary statistics for the main sample from 1986 to 2019. 
Panel A shows firm characteristics. Panel B shows forecast levels. Panel C shows forecast errors (realized minus predicted).Panel D shows forecast revisions (prediction at time $t$ minus that at time $t-1$).  Detailed variable definitions are provided in the Appendix.
\end{tablenotes}
\end{threeparttable}
\end{table}

\section{Evidence}
\label{sec:evidence}

This section provides our main empirical results. We start by showing that machine learning forecasts have accuracy that on average is fairly similar to human analysts. We then show that ML forecasts systematically violate forecast efficiency tests with patterns fairly similar to human analysts. To distinguish between behavioral and statistical explanations, we carry out three complementary identification strategies. These are experimental variation in regularization parameters, cross-sectional variation in signal quality, and quasi-experimental evidence from ML adoption timing. We conclude by providing alternative test specifications and considering the economic consequences for corporate investment.

\subsection{Forecast Accuracy}

Table \ref{tab:performance} compares forecasting performance across methods. Panel A gives the mean squared error (MSE) and mean absolute error (MAE) for the full sample and subperiods. Over the full 1986 to 2019 period, all three ML methods have lower MSE than human analysts. Random forests performs best with MSE of 0.892 compared to 0.959 for humans, roughly a 7\% improvement.

\begin{table}[htbp!]
\centering
\caption{\textbf{Forecast Performance Comparison}}
\label{tab:performance}
\begin{tabular}{lcccccc}
\toprule
\multicolumn{7}{l}{\textit{Panel A: Absolute Performance}} \\
\midrule 
 & \multicolumn{2}{c}{Full Sample} & \multicolumn{2}{c}{Pre-2013} & \multicolumn{2}{c}{Post-2013} \\
\cline{2-3} \cline{4-5} \cline{6-7}
Forecast Method & MSE & MAE & MSE & MAE & MSE & MAE \\
\midrule
Human Analysts & 0.959 & 0.487 & 1.004 & 0.499 & 0.767 & 0.432 \\
Ridge & 0.918 & 0.498 & 0.952 & 0.510 & 0.776 & 0.443 \\
Gradient Boosting & 0.901 & 0.490 & 0.923 & 0.498 & 0.804 & 0.456 \\
Random Forest & 0.892 & 0.484 & 0.917 & 0.492 & 0.783 & 0.446 \\
\midrule 
\multicolumn{7}{l}{\textit{Panel B: Win Rates (Percent of Firm-Years)}} \\
\midrule
 & \multicolumn{2}{c}{Full Sample} & \multicolumn{2}{c}{Pre-2013} & \multicolumn{2}{c}{Post-2013} \\
\cline{2-3} \cline{4-5} \cline{6-7}
 & Ratio & & Ratio & & Ratio & \\
\midrule 
Ridge Beats Human & 0.479 & & 0.482 & & 0.467 & \\
Gradient Boosting Beats Human & 0.500 & & 0.511 & & 0.454 & \\
Random Forest Beats Human & 0.519 & & 0.526 & & 0.486 & \\
\bottomrule 
\end{tabular}
\begin{flushleft}
\footnotesize
This table compares forecast performance across different methods. Panel A reports Mean Squared Error (MSE) and Mean Absolute Error (MAE) for each forecasting method. Panel B reports the percentage of firm-years where each method achieves lower MSE. All errors are computed using out-of-sample predictions.
\end{flushleft}
\end{table}

The performance gap gets much smaller after 2013. Pre-2013, random forests has an 8.7\% lower MSE than humans (0.917 versus 1.004). Post-2013, the gap reverses. Humans have MSE of 0.767 compared to 0.783 for random forests. The improved performance by humans reflects the fact that many of the analysts increasingly adopted ML methods themselves. So their forecasts become increasingly grounded by the ML methods.

\begin{figure}[htbp!]
	\caption{ML Win Rates Over Times}
	\label{fig:mlwins}
	\begin{center}
    \subcaption*{Panel A}
	\begin{minipage}{0.55\textwidth}
	\centering
	\includegraphics[width=1.0\linewidth]{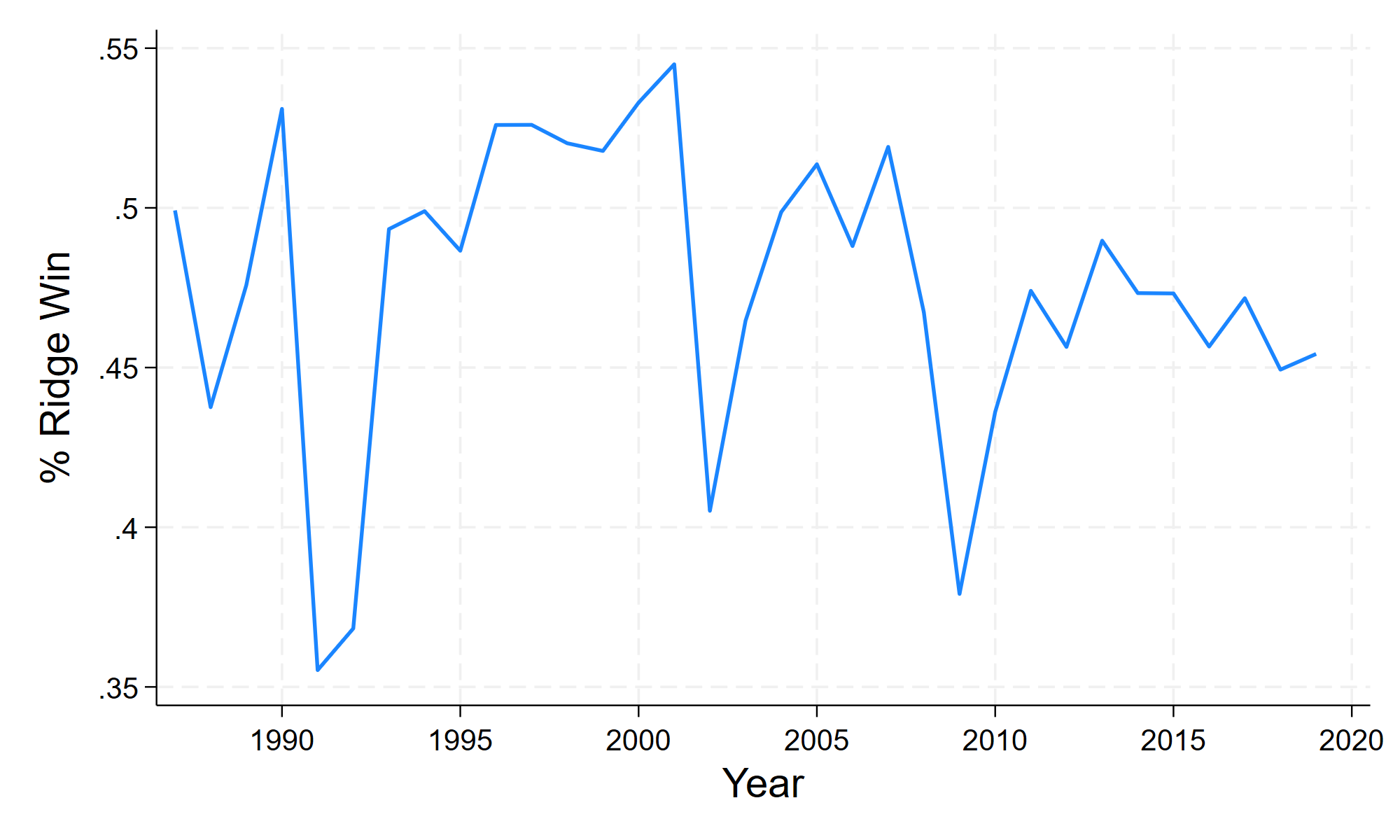}
	\end{minipage}
    \subcaption*{Panel B}
    \begin{minipage}{0.55\textwidth}
    \centering
    \includegraphics[width=1.0\linewidth]{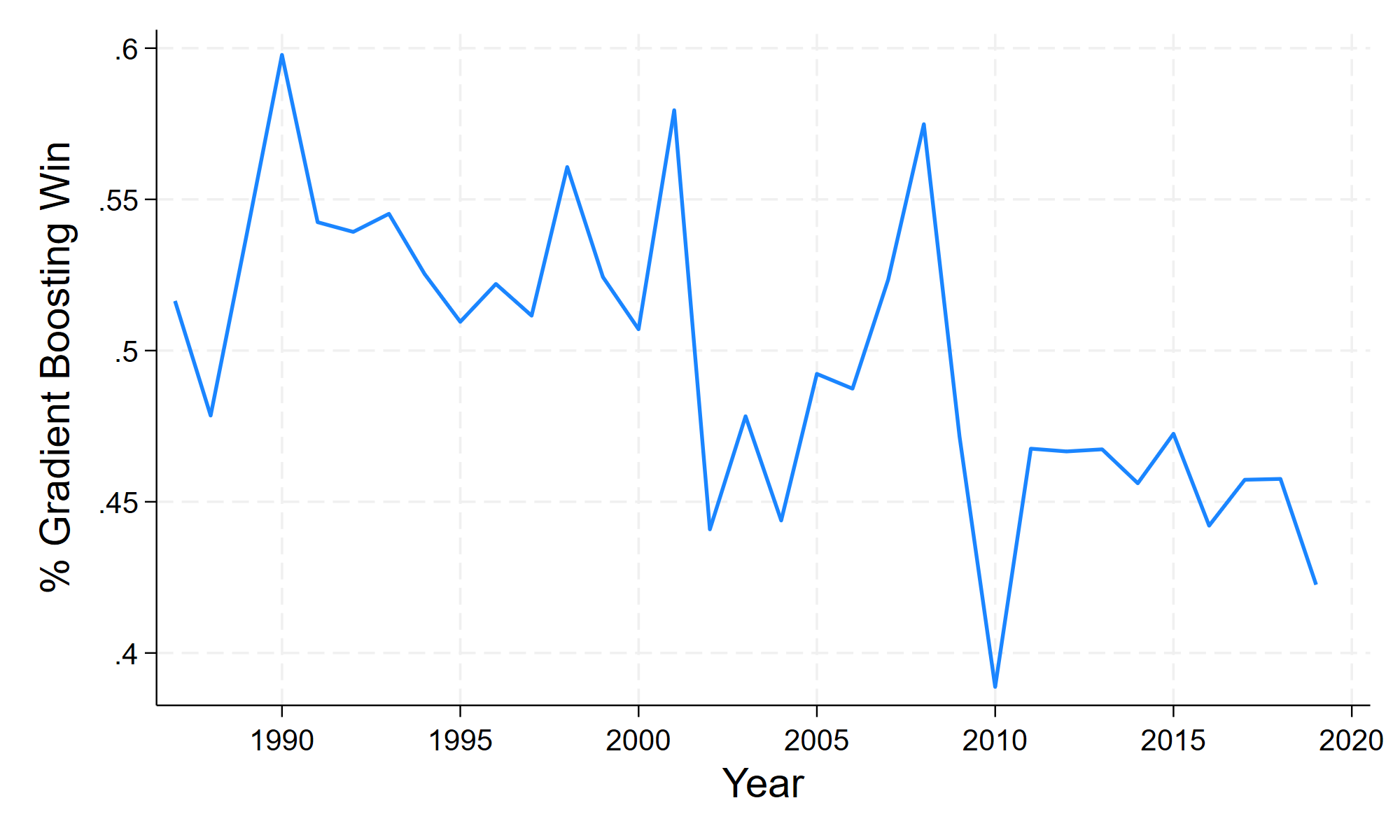}
    \end{minipage}
    \subcaption*{Panel C}
	\begin{minipage}{0.55\textwidth}
	\centering
	\includegraphics[width=1.0\linewidth]{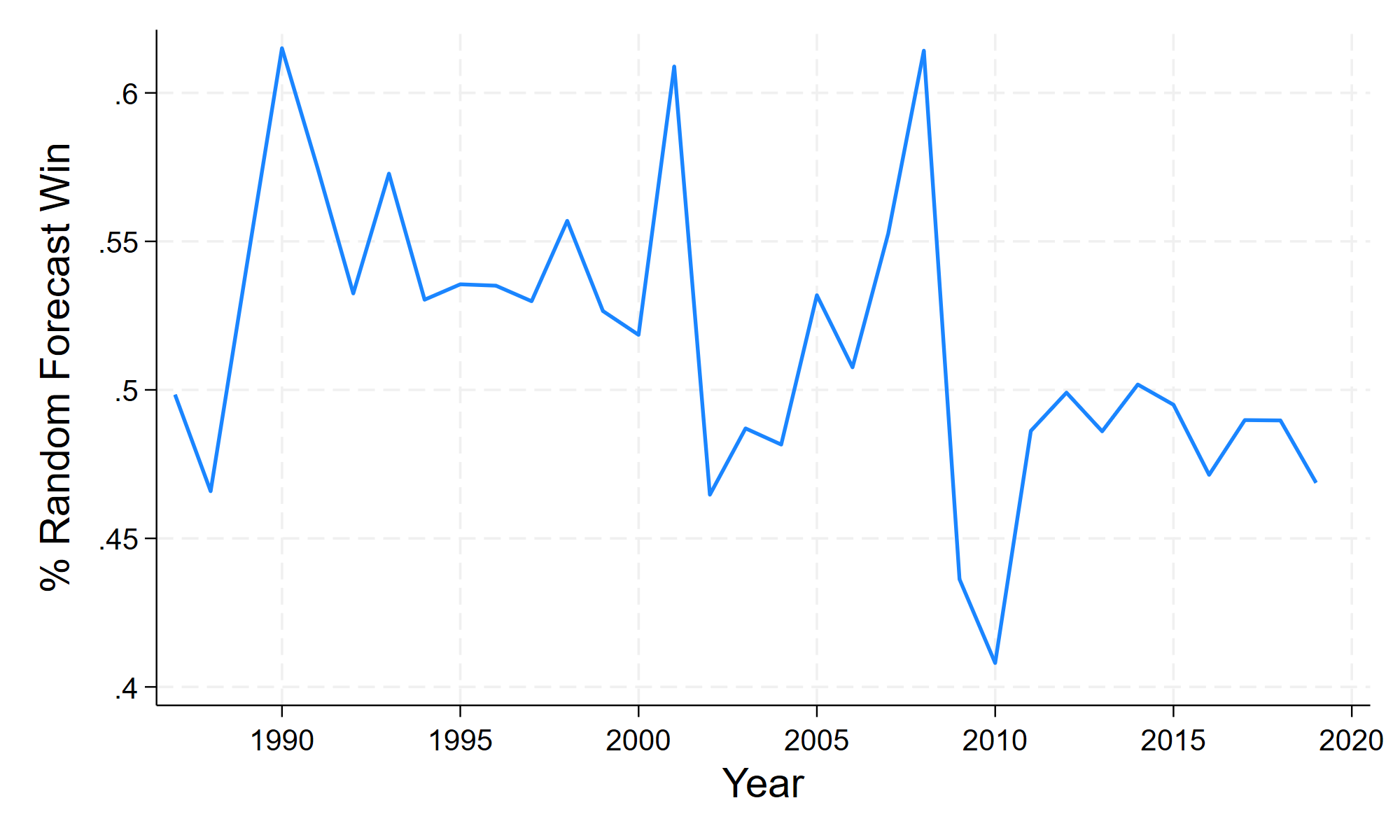}
	\end{minipage}
    \end{center}
    \begin{tablenotes}
    \small
    \item This figure shows the annual ratio of firm-year observations where ML outperforms human analysts. Panel A shows Ridge regression, Panel B shows Gradient Boosting, and Panel C shows Random Forest. 
    \end{tablenotes}
\end{figure}

Panel B shows the win rate. It is defined as the percentage of firm years in which each ML method produces more accurate forecasts than human analysts. Random forests beat humans 51.9\% of the time overall, with the advantage stronger pre-2013 (52.6\%) versus post-2013 (48.6\%). Figure \ref{fig:mlwins} plots win rates over time, showing the convergence pattern clearly.

These results establish that regularized forecasting methods achieve competitive accuracy, making it rational for forecasters to adopt such methods. The post-2013 convergence provides context for our quasi-experimental tests later in Section \ref{sec:mladoption}.

\clearpage

\subsection{Forecast Efficiency Violations}

Here we test whether ML forecasts satisfy the \citet{coibion2015information} forecast efficiency condition. Under rational expectations with perfect information and no regularization, forecast revisions should be uncorrelated with subsequent forecast errors. We estimate equation \ref{eq:CG_test}. All specifications include firm and year fixed effects with standard errors clustered by firm.

Table \ref{tab:cg_oneyear} shows the results for one year ahead forecasts. Column (1) shows human analysts have a significant positive coefficient of 0.114 (t-statistic 14.04). This pattern is commonly interpreted in the literature as underreaction. It says that forecasters do not revise enough when new information arrives.

Columns (2) through (4) apply the same test to forecasts produced by ML algorithms. Ridge regression has a coefficient of 0.006 (t-statistic 1.72) which is nearly zero and barely significant. Gradient boosting has  0.058 (t-statistic 6.37). Random forests has 0.044 (t-statistic 5.36). All ML methods exhibit significantly smaller coefficients than human analysts. The ridge predictions even approach the rational expectations benchmark of zero.

The economic magnitudes are large. A one unit increase in human forecast revision predicts an 11.4 percentage point larger forecast error. For gradient boosting, the same revision predicts only a 5.8 percentage point larger error, a 49\% reduction. The out of sample MSE at the bottom of Table \ref{tab:cg_oneyear} shows these efficiency patterns are not artifacts of overfitting. Ridge, gradient boosting, and random forests produce MSE of 0.918, 0.901, and 0.892 respectively. All of these out of sample evaluations are below human analysts' 0.959.

\begin{table}[htbp!]
\centering
\caption{\textbf{Machine Learning Forecast Errors Are Predictable: One-Year Horizon}}
\label{tab:cg_oneyear}
\begin{tabular}{lcccc}
\toprule
 & (1) & (2) & (3) & (4) \\
 & Human & Ridge & Gradient & Random \\
 & Analysts & & Boosting & Forest \\
\midrule 
 Dependent variable: & \multicolumn{4}{c}{$y_{i,t+1} - F_t y_{i,t+1}$} \\
\midrule
$F_t y_{i,t+1} - F_{t-1} y_{i,t+1}$ & 0.114*** & 0.006* & 0.058*** & 0.044*** \\
 & (14.04) & (1.72) & (6.37) & (5.36) \\
\midrule
Firm FE & Yes & Yes & Yes & Yes \\
Year FE & Yes & Yes & Yes & Yes \\
N & 99,963 & 99,963 & 99,963 & 99,963 \\
Adj $R^2$ & 0.16 & 0.11 & 0.11 & 0.10 \\
Out-of-sample MSE & 0.959 & 0.918 & 0.901 & 0.892 \\
\bottomrule
\end{tabular}
\begin{flushleft}
\footnotesize
This table reports OLS regression results testing whether forecast errors are predictable using forecast revisions, following \citet{coibion2015information}. The dependent variable is the one-year-ahead forecast error, defined as realized minus one-year-ahead predicted EPS. The main independent variable is the change in forecasts from $t-1$ to $t$. Standard errors (in parentheses) are clustered by firm. *, **, and *** indicate significance at the 10\%, 5\%, and 1\% levels, respectively.
\end{flushleft}
\end{table}

Table \ref{tab:cg_twoyear} evaluates two year ahead forecasts at a longer forecasting horizon. Human analysts have a near zero coefficient ($-0.003$, t-statistic $-0.24$). These forecasts are approximately efficient at longer horizons under standard tests. However, all three ML methods now produce large negative coefficients. Ridge is $-0.606$ (t-statistic $-27.41$), gradient boosting $-0.233$ (t-statistic $-19.05$), and random forests $-0.237$ (t-statistic $-19.68$).

\begin{table}[htbp!]
\centering
\caption{\textbf{Overreaction Patterns at Two-Year Horizon}}
\label{tab:cg_twoyear}
\begin{tabular}{lcccc}
\toprule
 & (1) & (2) & (3) & (4) \\
 & Human & Ridge & Gradient & Random \\
 & Analysts & & Boosting & Forest \\
\midrule
Dependent variable: & \multicolumn{4}{c}{$y_{i,t+2} - F_t y_{i,t+2}$} \\ 
\midrule
$F_t y_{i,t+2} - F_{t-1} y_{i,t+2}$ & $-0.003$ & $-0.606$*** & $-0.233$*** & $-0.237$*** \\
 & ($-0.24$) & ($-27.41$) & ($-19.05$) & ($-19.68$) \\
\midrule
Firm FE & Yes & Yes & Yes & Yes \\
Year FE & Yes & Yes & Yes & Yes \\
N & 58,864 & 58,864 & 58,864 & 58,864 \\
Adj $R^2$ & 0.12 & 0.36 & 0.18 & 0.18 \\
Out-of-sample MSE & 1.988 & 4.011 & 2.088 & 2.076 \\
\bottomrule
\end{tabular}
\begin{flushleft}
\footnotesize
This table reports OLS regression results testing whether forecast errors are predictable using forecast revisions, following \citet{coibion2015information}. The dependent variable is the two-year-ahead forecast error, defined as realized minus two-year-ahead predicted EPS. The main independent variable is the change in forecasts from $t-1$ to $t$. Standard errors (in parentheses) are clustered by firm. *, **, and *** indicate significance at the 10\%, 5\%, and 1\% levels, respectively.
\end{flushleft}
\end{table}
These negative coefficients imply overreaction. The upward forecast revisions predict negative forecast errors. Compared to human analysts, machine learning predictions produced much stronger overreaction at the two year horizon. There is a systematic sign change from positive at the short horizon (one year) to negative at the longer horizon (two years). This is consistent with the comparative statics results presented in Section \ref{sec:comp}. At short horizons (one year), the regularization shrinks coefficients toward zero, underweighting signals. This creates a positive force, i.e. underreaction to fundamentals. However, there is also the negative force which is the overreaction due to noise. At the short horizon these roughly balance out, creating nearly zero or slightly positive coefficients. In contrast, at longer horizons, the relationship between current signals and future earnings becomes more complex, introducing greater forecast noise with higher volatility as supported by \citet{de2024noise}. In this case, the noise effect dominates, resulting in negative CG coefficients.

The size of the differences across methods is noteworthy. The ridge regression coefficient of $-0.606$ is more than 2.5 times larger in absolute value than gradient boosting's $-0.233$. This presumably reflects ridge's linear structure. It applies uniform shrinkage without the nonlinear adaptivity of tree-based methods. Gradient boosting and random forests can learn different relationships at different horizons, partially compensating for mean reversion.

The out of sample MSE figures show that there is an important tradeoff. The forecast made by ridge has an MSE of 4.011 at the two-year horizon, nearly twice the error of human analysts (1.988). Gradient boosting (2.088) and random forests (2.076) perform fairly similarly to humans. This suggests ridge's strong negative coefficient reflects its regularization approach at longer horizons rather than superior forecasting.

These horizon effects are consistent with the findings documented in \citet{de2024noise}. There is a shift from small positive (nearly zero) coefficients at a one year horizon to large negative coefficients at the two year horizon. This is consistent with the predictions of noise in forecasts producing to systematic violations. Our contribution is to show that this pattern emerges from optimal regularization rather than requiring behavioral explanations.

\subsection{Robustness: Alternative Information Set}
To ensure that our results are not driven by a specific set of predictors, we compare forecast-efficiency violations using an alternative information set.

Table \ref{tab:cg_robusthorizons} examines whether our results are robust to an alternative information set by using a different set of predictors to generate the machine learning forecasts. Columns (1), (2), and (3) report one-year-ahead results, while Columns (4), (5), and (6) present two-year-ahead results. In contrast to Tables \ref{tab:cg_oneyear} and \ref{tab:cg_twoyear}, which use the full predictor set, these specifications rely on a more limited set of predictors that excludes WRDS financial ratios and employs default parameter settings.

\begin{table}[htbp!]
\centering
\caption{\textbf{Robustness of the Overreaction and Underreaction}}
\label{tab:cg_robusthorizons}
\begin{threeparttable}
\small
\begin{tabular}{lcccccc}
\toprule
& (1) & (2) & (3) & (4) & (5) & (6) \\
& Ridge  & Gradient & Random & Ridge  & Gradient & Random \\
&        & Boosting & Forest &        & Boosting & Forest \\
\cmidrule(lr){2-4} \cmidrule(lr){5-7}
& \multicolumn{3}{c}{$y_{i,t+1}   -F_{t} y_{i,t+1}$} & \multicolumn{3}{c}{$y_{i,t+2}   -F_{t} y_{i,t+2}$} \\
\midrule
$F_{t}   y_{i,t+1}   -F_{t-1} y_{i,t+1}$ & 0.049*** & 0.082*** & 0.042*** &           &           &           \\
& (6.74)   & (11.58)  & (5.56)   &           &           &           \\
$F_{t} y_{i,t+2}   -F_{t-1} y_{i,t+2}$   &          &          &          & -0.190*** & -0.227*** & -0.248*** \\
    &     &     &     & (-17.35)  & (-18.53)  & (-21.18)  \\
\midrule
Predictors & Limited  &  Limited  & Limited  &  Limited &  Limited  &  Limited  \\
Firm FE & Yes & Yes & Yes & Yes & Yes & Yes \\
Year FE & Yes & Yes & Yes & Yes & Yes & Yes \\
\midrule
N          & 99963    & 99963  & 90681    & 99963  & 99963    & 99963    \\
Adj R$^2$  & 0.13     & 0.11   & 0.10     & 0.09   & 0.11     & 0.10 \\
Out-of-sample MSE &	0.921	&	0.918	&	0.794	&	0.906	&	0.899	&	0.890 \\
\bottomrule
\end{tabular}
\begin{tablenotes}
\small
\item This table presents OLS regression results testing whether forecast errors are predictable using forecast revisions, following \citet{coibion2015information}. The dependent variable in Columns (1), (2), and (3) is the one-year-ahead forecast error, defined as realized minus predicted EPS. In Columns (4), (5), and (6), the dependent variable is the two-year-ahead forecast error, defined as realized minus predicted EPS. The main independent variable is the change in forecasts from t-1 to t. `Limited' indicates that financial ratios from the WRDS suite are excluded from the set of predictors. Standard errors (in parentheses) are clustered by firm. *, **, and *** indicate significance at the 10\%, 5\%, and 1\% levels, respectively. 
\end{tablenotes}
\end{threeparttable}
\end{table}

At the one-year horizon, limiting the predictor set has modest effects on CG coefficients. Ridge remains stable across specifications, with coefficients of 0.006 under the full predictor set and 0.049 under the limited predictor set. Gradient boosting shows a similar pattern: 0.058 with full predictors versus 0.082 with limited predictors. Random forests with full predictors yields 0.044, while the limited specification produces 0.042. The results are consistent with these algorithms' ability to internally select relevant features.

These patterns extend to longer horizons, and the predictions continue to exhibit overreacting forecasts. For ridge regression, the CG coefficient increases from –0.606 to –0.190. The estimates for gradient boosting change only marginally, from –0.233 to –0.227, while random forests range from –0.237 to –0.248. These findings suggest that the observed forecast efficiency tests violations are structural features inherent to the machine learning models rather than spurious artifacts driven by specific specifications.

\subsection{Experimental Evidence: Varying Regularization Intensity}

Tables \ref{tab:cg_oneyear} and \ref{tab:cg_twoyear} illustrate clearly that ML forecasts violate forecast efficiency tests. These violations may reflect behavioral biases rather than the effects of optimal regularization. To more clearly identify the impact of regularization, we perform simulation experiments that systematically vary the regularization parameters. Our theory predicts that stronger regularization should increase the CG coefficient (Proposition \ref{hyp:regularization}), shifting forecasts from overreaction toward underreaction.

\begin{table}[htbp!]
\centering
\caption{\textbf{Regularization Intensity and Forecast Bias}}
\label{tab:regularization}
\begin{tabular}{lcccccc}
\toprule
\multicolumn{7}{l}{\textit{Panel A: Ridge Penalty Alpha}} \\
Dependent variable: &\multicolumn{6}{c}{$y_{i,t+1} - F_t y_{i,t+1}$} \\
\midrule
 & (1) & (2) & (3) & (4) & (5) & (6) \\
 & 0.1 & 1 & 10 & 20 & 30 & 40 \\
 & (Low Reg) & & & & & (High Reg) \\
\midrule
$F_t y_{i,t+1} - F_{t-1} y_{i,t+1}$ & 0.001 & 0.002 & 0.004 & 0.005 & 0.006* & 0.006* \\
 & (0.24) & (0.61) & (1.21) & (1.50) & (1.72) & (1.89) \\
\midrule
Firm FE & Yes & Yes & Yes & Yes & Yes & Yes \\
Year FE & Yes & Yes & Yes & Yes & Yes & Yes \\
N & 99,963 & 99,963 & 99,963 & 99,963 & 99,963 & 99,963 \\
Adj $R^2$ & 0.11 & 0.11 & 0.11 & 0.11 & 0.11 & 0.11 \\
Out-of-Sample MSE & 0.928 & 0.925 & 0.921 & 0.920 & 0.918 & 0.917 \\
\midrule
\multicolumn{7}{l}{\textit{Panel B: Gradient Boosting Learning Rate}} \\
Dependent variable: &\multicolumn{6}{c}{$y_{i,t+1} - F_t y_{i,t+1}$} \\
\midrule
 & (1) & (2) & (3) & (4) & (5) & (6) \\
 & 0.01 & 0.05 & 0.1 & 0.2 & 0.3 & 0.4 \\
 & (High Reg) & & & & & (Low Reg) \\
\midrule 
$F_t y_{i,t+1} - F_{t-1} y_{i,t+1}$ & 0.675*** & 0.129*** & 0.058*** & 0.015* & $-0.012$ & $-0.040$*** \\
 & (30.42) & (12.99) & (6.37) & (1.65) & ($-1.37$) & ($-4.65$) \\
\midrule
Firm FE & Yes & Yes & Yes & Yes & Yes & Yes \\
Year FE & Yes & Yes & Yes & Yes & Yes & Yes \\
N & 99,963 & 99,963 & 99,963 & 99,963 & 99,963 & 99,963 \\
Adj $R^2$ & 0.48 & 0.17 & 0.11 & 0.09 & 0.09 & 0.09 \\
Out-of-Sample MSE & 2.242 & 0.962 & 0.901 & 0.907 & 0.923 & 0.946 \\
\midrule 
\multicolumn{7}{l}{\textit{Panel C: Formal Tests}} \\
\midrule
 & \multicolumn{3}{c}{(1)} & \multicolumn{3}{c}{(2)} \\
 & \multicolumn{3}{c}{Ridge} & \multicolumn{3}{c}{Gradient Boosting} \\
\midrule 
& \multicolumn{6}{c}{CG Coefficient} \\
\midrule 
Regularization & \multicolumn{3}{c}{0.00009***} & \multicolumn{3}{c}{0.008***} \\
 & \multicolumn{3}{c}{(16.13)} & \multicolumn{3}{c}{(5.87)} \\
\midrule
N & \multicolumn{3}{c}{41} & \multicolumn{3}{c}{40} \\
Adj $R^2$ & \multicolumn{3}{c}{0.87} & \multicolumn{3}{c}{0.46} \\
\bottomrule 
\end{tabular}
\begin{flushleft}
\footnotesize
This table reports OLS regression results examining whether the predictability of forecast errors is related to the degree of regularization in machine learning models. Panels A and B report results testing whether forecast errors are predictable using forecast revisions, following \citet{coibion2015information}. The dependent variable is the one-year-ahead forecast error. In Panel A, the hyperparameter varied in Ridge Regression is the ridge penalty $\alpha$, ranging from 0.1 to 40. In Panel B, the hyperparameter varied in Gradient Boosting is the learning rate, ranging from 0.01 to 0.4. Panel C regresses the CG coefficient on the ranked order of regularization intensity. Standard errors (in parentheses) are clustered by firm. *, **, and *** indicate significance at the 10\%, 5\%, and 1\% levels, respectively.
\end{flushleft}
\end{table}

Table \ref{tab:regularization} gives the evidence. Panel A varies ridge regression's penalty parameter $\alpha$ from 0.1 (low regularization) to 40 (high regularization). The CG coefficient increases monotonically from 0.001 to 0.006. It goes from statistically insignificant to significant (t-statistic 1.89) as regularization increases. The magnitude is modest in levels, the monotonic relationship strongly supports our mechanism.

Panel B considers the gradient boosting learning rate. A lower values means stronger regularization. The CG coefficient decreases dramatically from $0.675$ to $-0.040$ as the learning rate increases from 0.01 to 0.4. At very low learning rates, gradient boosting produces large positive CG coefficients, showing strong underreaction. At higher learning rates with almost no regularization the coefficient turns negative. This is consistent with overfitting to transient noise fluctuations.

The out of sample MSE has a U-shape. The minimum is at learning rate 0.1. That is where gradient boosting achieves its optimal bias-variance tradeoff. Notice that the CG coefficient at this optimum is 0.058. This reflects the fact that optimal regularization produces forecast efficiency violations. This reflects our main theoretical prediction. Rational statistical practice systematically fails standard efficiency tests.

The CG coefficient can be made to become insignificant. In Table \ref{tab:regularization}, that happens at a learning rate of 0.30. But getting rid of that bias comes at a cost. In that case the out of sample MSE is worse than with somewhat higher regularization. This is an illustration of the fundamental tradeoff. Forecasters must in effect choose between satisfying in sample efficiency tests and minimizing out of sample prediction error. Optimal statistical practice quite reasonably focuses on out of sample performance. 

Panel C formalizes this relationship by regressing estimated CG coefficients on ranked regularization intensity. For ridge regression, we estimate 41 models varying $\alpha$ from 0.1 to 40. We rank them by regularization intensity, and then regress each model's CG coefficient on its rank. There is a strong positive relationship, with a coefficient of 0.00009 which is positive and statistically significant. For gradient boosting, we vary learning rates from 0.01 to 0.4 across 40 models. The coefficient is 0.008, positive and statistically significant. This again  confirms the relationship.

\begin{figure}[htbp!]
\centering
\caption{\textbf{Regularization Intensity and CG Coefficients}}
\label{fig:regularization}
\begin{center}
	\subcaption*{Panel A: Ridge Regression}
	\begin{minipage}{0.8\textwidth}
	\centering
	\includegraphics[width=1.0\linewidth]{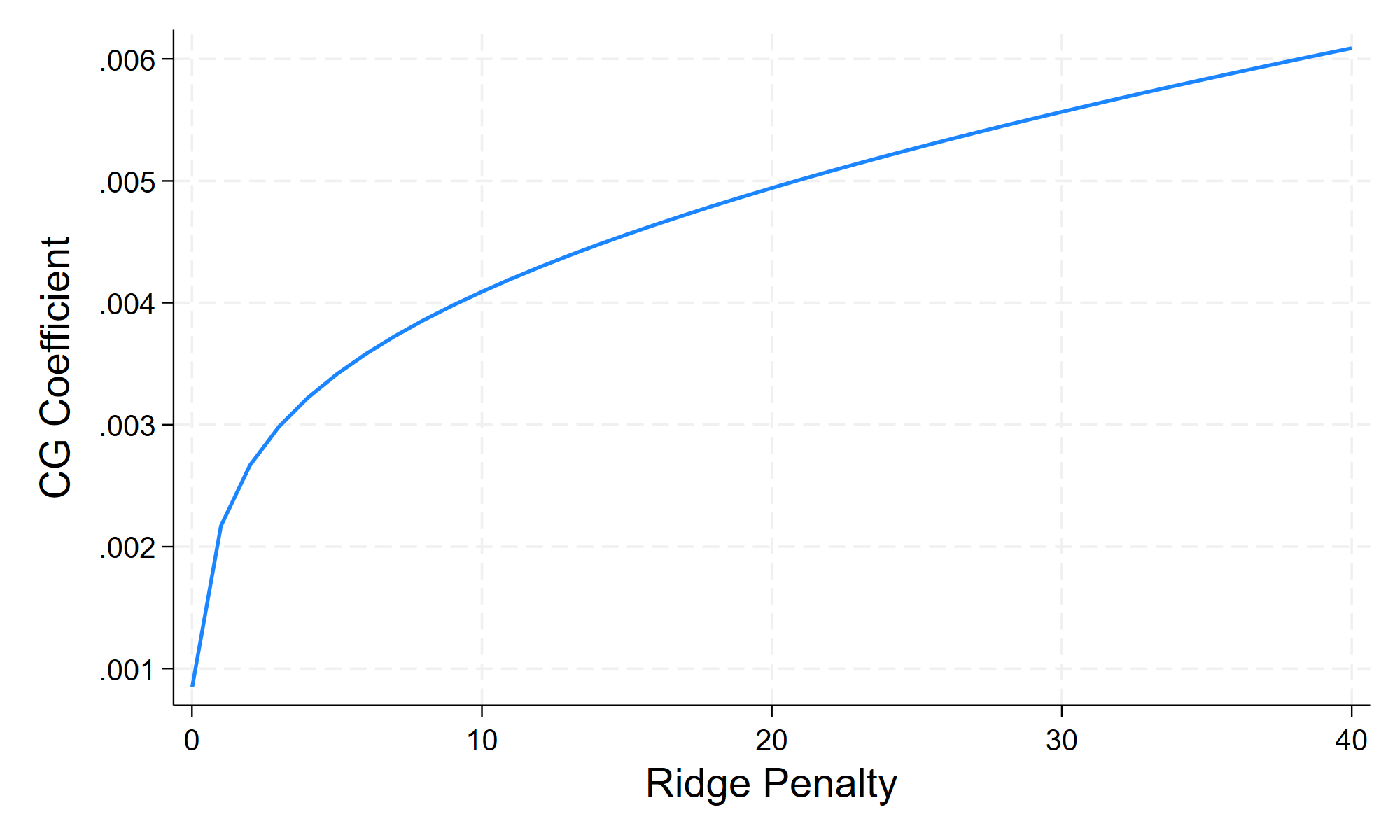}
	\end{minipage}

    \subcaption*{Panel B: Gradient Boosting}
    \begin{minipage}{0.8\textwidth}
    \centering
    \includegraphics[width=1.0\linewidth]{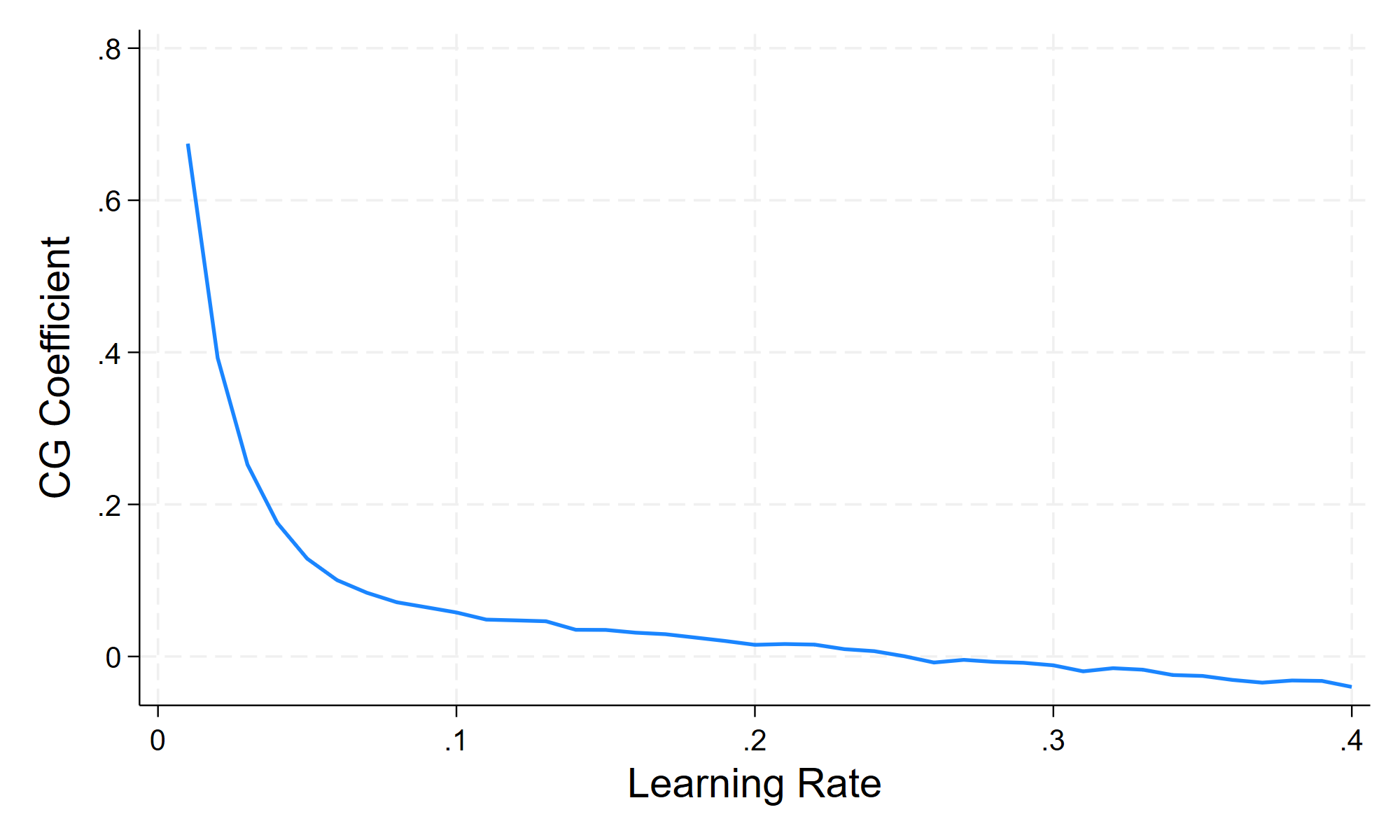}
    \end{minipage}
\end{center}
\begin{flushleft}
\footnotesize
This figure plots the \citet{coibion2015information} coefficient against ML predictions under different hyperparameter settings. In Panel A, the hyperparameter varied in Ridge Regression is the ridge penalty $\alpha$, ranging from 0.1 to 40. In Panel B, the hyperparameter varied in Gradient Boosting is the learning rate, ranging from 0.01 to 0.4.
\end{flushleft}
\end{figure}

Figure \ref{fig:regularization} provides a visual illustration of these relationships. Panel A plots CG coefficients against ridge penalties, showing the gradual increase from near zero to 0.006. Panel B plots coefficients against gradient boosting learning rates. This shows the dramatic decline from 0.68 to negative values. The nonlinear pattern in Panel B reflects gradient boosting's complex optimization, but the monotonic relationship is clear.

These results provide experimental evidence of our mechanism using simulations. When exogenously varying the regularization intensity while holding all else constant, we indeed observe the predicted effects on forecast rationality violations. This setup ensures clean identification. Later in Section \ref{sec:mladoption}, we extend this analysis to a quasi-natural experiment that examines how the adoption of ML methods by analysts, thus changes in the regularization of their forecasts, in the real world generates similar implications.

\subsection{Cross-Sectional Evidence: Signal Quality}

According to our theory, forecast efficiency violations vary systematically with noise persistence and volatility. Propositions \ref{hyp:noise_persistence} and \ref{hyp:signal_quality}  predict more negative CG coefficients when noise persistence $\rho_\eta$ is low, and when noise volatility $\sigma_\eta$ is high. Testing these cross-sectional predictions is helpful for distinguishing between rational regularization and behavioral biases. Behavioral explanations are based on the human psychology of the analyst. So they normally predict uniform effects across all firms being covered by the analyst. In contrast, optimal regularization depends on firm characteristics. The mechanism predicts heterogeneity that reflects firm characteristics such as noise persistence and volatility.

To examine this, we interact forecast revisions with firm characteristics that proxy for signal quality. We use R\&D expenditure scaled by total assets and firm age. The idea is that higher R\&D intensity reflects uncertainty about future cash flows. Younger firms should have noisier signals. These characteristics capture lower noise persistence and higher noise volatility. The empirical specification is
\begin{equation}
e_{i,t+1} = \gamma_1 r_{i,t} + \gamma_2 r_{i,t} \times \text{Characteristic}_i + \alpha_i + \delta_t + \varepsilon_{i,t+1}.
\end{equation}
Table \ref{tab:crosssection} reports results. Panel A shows the effect of heterogeneity in R\&D expenditure. We split firms into two groups based on whether they are above or below the sample average of R\&D expenditure. For human analysts, Column (1) shows a CG coefficient of 0.125 for low R\&D firms compared to Column (2) showing 0.103 for high R\&D firms. This means that human analysts have stronger overreaction (a lower coefficient) for high R\&D firms.

The pattern is stronger in ML forecasts. For ridge regression, the CG coefficient is 0.033 for low R\&D firms and $-0.000$ for high R\&D firms. For gradient boosting, coefficients are 0.092 for low R\&D firms and 0.034 for high R\&D firms. Columns (3), (6), and (9) include interaction terms between forecast revisions and R\&D expenditure. The coefficients are $-0.538$, $-0.301$, and $-0.454$ respectively. All of them are negative and statistically significant. These findings confirm the prediction that firms with higher R\&D intensity have negative CG coefficients and stronger overreaction. This makes sense if high R\&D is a reasonable proxy for lower noise persistence and higher noise volatility.

Panel B uses firm age as another proxy for noise persistence and volatility. Younger firms are presumably more uncertain and therefore have lower noise persistence and higher noise volatility. For analyst predictions, the CG coefficient is 0.049 for younger firms and 0.139 for older firms. For ridge regression, the coefficient is $-0.034$ for younger firms and 0.011 for older firms. Noise can even flip the sign of the CG coefficient, amplifying the magnitude of the effect. For gradient boosting, coefficients are 0.026 for younger firms and 0.069 for older firms.

Columns (3), (6), and (9) include interaction terms between forecast revisions and firm age. The coefficients are 0.002 for all three forecasting methods, all positive and statistically significant. These findings show that as predicted older firms have less negative (or more positive) CG coefficients.

\begin{sidewaystable}[htbp!]
\centering
\caption{\textbf{Cross-sectional Heterogeneity}}
\label{tab:crosssection}
\begin{threeparttable}
\small
\begin{tabular}{lccccccccc}
\toprule
\multicolumn{10}{l}{\textit{Panel A: Heterogeneity in R\&D Expenditure}} \\
\midrule
& (1) & (2) & (3) & (4) & (5) & (6) & (7) & (8) & (9)\\
& \multicolumn{3}{c}{Analysts} & \multicolumn{3}{c}{Ridge} & \multicolumn{3}{c}{Gradient Boosting}\\
\cmidrule(lr){2-4}\cmidrule(lr){5-7}\cmidrule(lr){8-10}
& Low & High & & Low & High & & Low & High & \\
\cmidrule(lr){2-10}
& \multicolumn{9}{c}{$y_{i,t+1}   -F_{t} y_{i,t+1}$} \\
\midrule
$F_{t}   y_{i,t+1}   -F_{t-1} y_{i,t+1}$ & 0.125*** & 0.103*** & 0.123***  & 0.033*** & -0.000  & 0.030***  & 0.092*** & 0.034*** & 0.091***  \\
                                         & (10.06)  & (9.73)   & (13.14)   & (3.18)   & (-0.13) & (3.86)    & (8.72)   & (2.58)   & (11.15)   \\
$(F_{t} y_{i,t+1}   -F_{t-1} y_{i,t+1})$ &          &          & -0.538*** &          &         & -0.301*** &          &          & -0.454*** \\
$\times R\&D Expenditure$                &          &          & (-7.82)   &          &         & (-5.38)   &          &          & (-6.94)   \\
Firm FE                                  & Yes      & Yes      & Yes       & Yes      & Yes     & Yes       & Yes      & Yes      & Yes       \\
Year FE                                  & Yes      & Yes      & Yes       & Yes      & Yes     & Yes       & Yes      & Yes      & Yes       \\
N                                        & 40913    & 59050    & 81896     & 40913    & 59050   & 81896     & 40913    & 59050    & 81896     \\
Adj R$^2$                                & 0.15     & 0.16     & 0.14      & 0.11     & 0.11    & 0.10      & 0.12     & 0.11     & 0.10      \\
\midrule
\multicolumn{10}{l}{\textit{Panel B: Heterogeneity in Firm Age}} \\
\midrule
& (1) & (2) & (3) & (4) & (5) & (6) & (7) & (8) & (9)\\
& \multicolumn{3}{c}{Analysts} & \multicolumn{3}{c}{Ridge} & \multicolumn{3}{c}{Gradient Boosting}\\
\cmidrule(lr){2-4}\cmidrule(lr){5-7}\cmidrule(lr){8-10}
& Low & High & & Low & High & & Low & High & \\
\cmidrule(lr){2-10}
& \multicolumn{9}{c}{$y_{i,t+1}   -F_{t} y_{i,t+1}$} \\
\midrule
$F_{t}   y_{i,t+1}   -F_{t-1} y_{i,t+1}$       & 0.049***      & 0.139*** & 0.063*** & -0.034*** & 0.011*** & -0.028*** & 0.026** & 0.069*** & 0.033*** \\
                                               & (4.16)        & (13.51)  & (5.07)   & (-3.66)   & (3.10)   & (-2.84)   & (2.23)  & (5.83)   & (2.82)   \\
\multicolumn{2}{l}{$(F_{t}   y_{i,t+1}   -F_{t-1} y_{i,t+1})$} &          &          &           &          &           &         &          &          \\
$\times Age$                                   &               &          & 0.002*** &           &          & 0.002***  &         &          & 0.002*** \\
                                               &               &          & (3.79)   &           &          & (5.46)    &         &          & (4.24)   \\
Firm FE                                        & Yes           & Yes      & Yes      & Yes       & Yes      & Yes       & Yes     & Yes      & Yes      \\
Year FE                                        & Yes           & Yes      & Yes      & Yes       & Yes      & Yes       & Yes     & Yes      & Yes      \\
N                                              & 40464         & 59499    & 81899    & 40464     & 59499    & 81899     & 40464   & 59499    & 81899    \\
Adj R$^2$                                      & 0.16          & 0.17     & 0.14     & 0.12      & 0.12     & 0.10      & 0.12    & 0.12     & 0.10    \\
\bottomrule
\end{tabular}
\begin{tablenotes}
\small
\item 
This table reports OLS regression results testing whether the predictability of forecast errors, based on forecast revisions following \citet{coibion2015information}, differs by R\&D expenditure and firm age. In Panel A, the sample is split into two groups based on firms’ average R\&D expenditure. Columns (1), (4), and (7) include firms below the sample-average R\&D expenditure, while columns (2), (5), and (8) include firms above the average. Columns (3), (6), and (9) additionally include the interaction term between R\&D expenditure and forecast revisions. In Panel B, the sample is similarly split into two groups based on firms’ average age. *, **, *** indicate significance at 10\%, 5\%, and 1\% levels.
\end{tablenotes}
\end{threeparttable}
\end{sidewaystable}

\subsection{Natural Experiment Evidence: Machine Learning Adoption}\label{sec:mladoption}

The cross-section tests and the simulations provide strong evidence that regularization drives forecast efficiency violations. Here we use a third identification strategy based on the timing of ML adoption. Around the year 2013, ML methods became much more popular and more widely used by finance practitioners. Suppose that analysts began adopting regularization methods around 2013. Then forecast patterns should show a distinct shift around that time. Not all stock analysts are likely to make the change at the same rate. Analysts with quantitative training are expected to adopt ML methods earlier and more aggressively since they have the background training. They should show stronger changes in forecast patterns.

There are several reasons to identify 2013 as a critical inflection point in ML adoption. The popular open source python library scikit-learn reached its first stable version (0.10) in 2012. That made previously specialized statistical methods easily accessible to analysts without extensive programming expertise and efforts. That library includes ridge regression, lasso, elastic net, random forests, and gradient boosting with standardized APIs. This dramatically reduced the cost of implementing formal regularized forecasting.

We classify analysts as `technical' if they hold undergraduate or graduate degrees in statistics, mathematics, computer science, engineering, or related quantitative fields. We identified this training by hand collecting and verifying individual LinkedIn profiles and FINRA BrokerCheck records. Our sample includes 173 technical analysts and 685 non-technical analysts covering 14,901 and 36,155 firm-year observations respectively. For each firm-month, we calculate median consensus forecasts for technical analysts and non-technical analysts separately.
We do not know who actually used ML. But under the maintained hypothesis that technical training is a proxy for ML use, we can examine if the associated forecasts make sense. 

Table \ref{tab:adoption} shows whether analyst technical background helps explain forecast efficiency patterns before and after 2013. Columns (1) and (2) have CG regression estimates for the pre-2013 period. Technical analysts show a coefficient of 0.020 (t-statistic 0.43, insignificant), while non-technical analysts exhibit a coefficient of 0.107 (t-statistic 2.35, significant). Both coefficients are positive, consistent with mild underreaction during this period.
\begin{table}[htbp!]
\centering
\caption{\textbf{Analyst Technical Background and ML Adoption}}
\label{tab:adoption}
\begin{tabular}{lcccc}
\toprule
 & \multicolumn{2}{c}{Pre-2013} & \multicolumn{2}{c}{Post-2013} \\
 & (1) & (2) & (3) & (4) \\
 & Technical & Non-Tech & Technical & Non-Tech \\
\midrule 
Dependent variable: & & \multicolumn{2}{c}{$y_{i,t+1} - F_t y_{i,t+1}$} \\
\midrule
$F_t y_{i,t+1} - F_{t-1} y_{i,t+1}$ & 0.020 & 0.107** & $-0.147$*** & $-0.014$ \\
 & (0.43) & (2.35) & ($-3.90$) & ($-0.19$) \\
\midrule 
Firm FE & Yes & Yes & Yes & Yes \\
Year FE & Yes & Yes & Yes & Yes \\
N & 4,813 & 5,282 & 5,564 & 6,012 \\
Adj $R^2$ & 0.15 & 0.15 & 0.16 & 0.09 \\
\bottomrule 
\end{tabular}
\begin{flushleft}
\footnotesize
This table examines whether analyst technical background predicts differential adoption of ML methods. Technical analysts have undergraduate or graduate degrees in statistics, computer science, mathematics, engineering, or related quantitative fields, identified from LinkedIn and FINRA records. The sample spans 1994 to 2018 and is split at 2013, when ML tools became widely accessible through platforms such as scikit-learn. The dependent variable is the analyst forecast error. All regressions include firm and year fixed effects. Standard errors (in parentheses) are clustered by firm. *, **, *** indicate significance at 10\%, 5\%, and 1\% levels.
\end{flushleft}
\end{table}
There is a marked change between before and after 2013. This is shown in Columns (3) and (4). After 2013 technical analysts have a strongly negative coefficient of $-0.147$ (t-statistic $-3.90$), showing strong overreaction. Non-technical analysts also shift negative, but to a much smaller extent with a coefficient of $-0.014$ (t-statistic $-0.19$, insignificant). The difference is economically large and statistically significant.

Technical analysts show a decline of 0.167 in the CG coefficient (from 0.020 pre-2013 to $-0.147$ post-2013). In contrast, non-technical analysts have a much smaller decline of 0.121 (from 0.107 to $-0.014$). This different response strongly supports our hypothesis that technically trained analysts adopted ML methods more aggressively following the widespread availability of accessible tools. The shift toward overreaction is consistent with a reduction in effective regularization intensity. This is presumably because analysts moved from very strong implicit regularization that is characteristic of manual forecasting. They moved to more moderate explicit regularization that is characteristic of ML methods.

A reasonable concern is that technical and non-technical analysts may cover different types of firms. That potentially confounds our interpretation. The inclusion of firm fixed effects in all specifications is intended to address this by comparing forecasts for the same firm before and after 2013. This quasi-experimental design provides further evidence that statistical regularization, not psychology or changing firm characteristics, drives the observed forecast patterns.

\subsection{Robustness Evidence: Level Test}

The \citet{coibion2015information} test is the dominant testing method in the modern literature. But there are other tests that are used in some papers. \citet{bordalo2018diagnostic} use a simple levels test that regresses forecast errors on the level of signals rather than their changes.
\begin{equation}
e_{i,t+2} = \delta_B z_{i,t} + \alpha_i + \delta_t + \nu_{i,t},
\end{equation}
where $z_{i,t}$ represents salient signals. Under rational expectations, $\delta_B = 0$. A positive coefficient is interpreted as forecasters underweighting salient signals (underreaction). A negative coefficient is interpreted as overreaction. Table \ref{tab:levels_tests} carries out these tests using investment and net debt issuance as signals.

\begin{table}[htbp!]
\centering
\caption{\textbf{Levels Tests for Overreaction}}
\label{tab:levels_tests}
\begin{tabular}{lcccc}
\toprule
\multicolumn{5}{l}{\textit{Panel A: Investment as Signal}} \\
\midrule 
 & (1) & (2) & (3) & (4) \\
 & Human & Ridge & Gradient & Random \\
 & Analysts & & Boosting & Forest \\
\midrule
\multicolumn{5}{c}{$y_{i,t+2} - F_t y_{i,t+2}$} \\
\midrule
Investment & $-1.860$*** & $-2.285$*** & $-2.061$*** & $-2.154$*** \\
 & ($-13.947$) & ($-16.598$) & ($-15.715$) & ($-16.236$) \\
\midrule
Firm FE & Yes & Yes & Yes & Yes \\
Year FE & Yes & Yes & Yes & Yes \\
N & 83,084 & 83,084 & 83,084 & 83,084 \\
Adj $R^2$ & 0.14 & 0.12 & 0.15 & 0.14 \\
\midrule
\multicolumn{5}{l}{\textit{Panel B: Debt Issuance as Signal}} \\
\midrule
 & (1) & (2) & (3) & (4) \\
 & Human & Ridge & Gradient & Random \\
 & Analysts & & Boosting & Forest \\
\midrule
\multicolumn{5}{c}{$y_{i,t+2} - F_t y_{i,t+2}$} \\
\midrule
Debt Net Issuance & $-0.729$*** & $-0.783$*** & $-0.785$*** & $-0.800$*** \\
 & ($-17.089$) & ($-17.486$) & ($-17.922$) & ($-18.306$) \\
\midrule
Firm FE & Yes & Yes & Yes & Yes \\
Year FE & Yes & Yes & Yes & Yes \\
N & 83,672 & 83,672 & 83,672 & 83,672 \\
Adj $R^2$ & 0.14 & 0.11 & 0.15 & 0.13 \\
\bottomrule 
\end{tabular}
\begin{flushleft}
\footnotesize
This table reports OLS regression results testing whether forecast errors are predictable using the levels of signals, following \citet{bordalo2018diagnostic}. The dependent variable is the two-year-ahead forecast error. In Panel A, the signal is investment rate. In Panel B, the signal is net debt issuance. Standard errors (in parentheses) are clustered by firm. *, **, and *** indicate significance at the 10\%, 5\%, and 1\% levels, respectively.
\end{flushleft}
\end{table}
In Table \ref{tab:levels_tests} panel A uses investment rate as the signal. Column (1) shows human forecasts have a coefficient of $-1.860$ (t-statistic $-13.947$). This confirms the widely documented finding that human forecasts overreact at longer horizons. Columns (2) to (4) show ML forecasts have even stronger overreaction, with coefficients of $-2.285$, $-2.061$, and $-2.154$ respectively. All of them are highly significant.

Panel B uses net debt issuance as another news signal. The results are similar. Human forecasts have a coefficient of $-0.729$. ML forecasts range from $-0.783$ to $-0.800$. These findings confirm that ML forecasts also exhibit overreaction at longer horizons. It is based on a different test specification that is used in the literature. This reinforces the evidence for  our main results.


\subsection{Economic Consequences}

Our analysis so far shows that regularization produces systematic forecast efficiency violations. We now consider whether these patterns have economic consequences for corporate investment decisions. Suppose that managers respond to analyst forecasts when allocating capital, and suppose that those forecasts contain systematic biases from regularization. Then we should observe correlations between predicted forecast errors and subsequent investment changes.

To see if that happens, we follow the two step approach used by \citet{bordalo2025real}. The first step is to regress forecast errors on forecast revisions. This is intended to isolate the component of forecast errors that reflects regularization induced bias. The second step is to see if these predicted forecast errors correlate with subsequent investment changes. This approach permits us to get the part of the forecast errors that are due to biased use of information; at the same time, it controls for other unobservable factors that affect both forecast errors and investment.

We study the post-2013 sample since that is when machine learning tools became widely accessible. This period is suitable for distinguishing between technical analysts who adopted ML methods more aggressively, and non-technical analysts who did not. We restrict the second-stage sample to firms for which more than 40\% of covering analysts have traceable educational backgrounds. This is intended to allow adequate representation of both analyst types without losing an excessive number of observations.

Table \ref{tab:investment} presents the results. Columns (1) and (2) show the first-stage regressions. These just replicate the findings in Table \ref{tab:adoption}. For technical analysts post-2013, the CG coefficient is $-0.147$ (t-statistic $-3.90$), showing strong overreaction. For non-technical analysts, the coefficient is $-0.014$ (t-statistic $-0.19$) and it is statistically insignificant. These first-stage results show that technical analysts exhibit stronger forecast efficiency violations once ML methods are adopted.

\begin{table}[htbp!]
\centering
\caption{\textbf{Economic Consequences: Forecast Errors and Investment Reversals}}
\label{tab:investment}
\begin{threeparttable}
\small
\begin{tabular}{lcccc}
\toprule
& (1) & (2) & (3) & (4) \\
& Tech & NonTech & Tech & NonTech \\
\cmidrule(lr){2-3} \cmidrule(lr){4-5}
Dependent Variable: & \multicolumn{2}{c}{ $y_{i,t+1} - F_t y_{i,t+1}$ } & \multicolumn{2}{c}{ $\Delta$Investment$_{t,t+1}$} \\
\midrule
Predicted $y_{i,t+1} - F_t y_{i,t+1}$ &  &     & 0.007**  & 0.006**  \\
 &  &     & (2.21)  & (2.42)  \\
$F_t y_{i,t+1} - F_{t-1} y_{i,t+1}$  & -0.147***  & -0.014   &  &           \\
                   & (-3.90)   & (-0.19) &  &           \\
Profitability$_{i,t}$ &  &    & 0.021*** & 0.028*** \\
 &  &    & (4.57)  & (8.42)  \\

\midrule
Firm FE & Yes & Yes & No & No  \\
Year FE & Yes & Yes & Yes & Yes  \\
N      & 5,564   & 6,012   & 1,914 & 2,076   \\
Adj R$^2$  & 0.16     & 0.09   & 0.05   & 0.06  \\    
\bottomrule
\end{tabular}
\begin{tablenotes}
\small
\item This table examines whether forecast errors predict subsequent changes in corporate investment from 2013 to 2018. Columns (1) and (2) report the IV first-stage estimates, regressing forecast errors on investment. The dependent variable is the forecast error, $y_{i,t+1} - F_{t} y_{i,t+1}$, and the independent variable is forecast revision at time $t$, $F_t y_{i,t+1} - F_{t-1} y_{i,t+1}$. Columns (3) and (4) report the second-stage regression, showing how instrumented forecast errors affect investment. The dependent variable is the change in investment rate from year $t$ to year $t+1$. Predicted forecast errors are estimated from first stage regressions of forecast errors on forecast revisions. We restrict the second stage sample to firms with more than 40\% of the analysts for whom we can trace educational background. The mean value of investment rate change is 0.034. The standard deviations of predicted forecast errors are 0.50 and 0.47 for tech and non-tech analysts, respectively. Standard errors (in parentheses) are clustered by firm. *, **, *** indicate significance at 10\%, 5\%, and 1\% levels.
\end{tablenotes}
\end{threeparttable}
\end{table}
Columns (3) and (4) provide the second stage results. The coefficient on predicted forecast errors is 0.007 (t-statistic 2.21) for technical analysts, and 0.006 (t-statistic 2.42) for non-technical analysts. Both coefficients are positive and statistically significant, which implies that the predicted forecast errors are associated with subsequent investment changes in the same direction. 

These effects are large enough to be economically significant. The mean value of investment rate change is 0.034. The standard deviations of predicted forecast errors are 0.50 and 0.47 for tech and non-tech analysts, respectively. For tech analysts, a one–standard deviation increase in predicted forecast error leads to a $10\% (= 0.007 \times 0.50 / 0.034)$ increase in investment changes. For non-tech analysts, a one–standard deviation increase in predicted forecast error leads to an $8\% (= 0.006 \times 0.47 / 0.034)$ increase in investment changes.

This shows that technical analysts post-2013 exhibit overreaction (negative CG coefficients). Due to regularization, these revisions overshoot. This results in forecast errors such that actual earnings fall short of predictions. The positive second stage coefficient of 0.007 means that these negative forecast errors after positive revisions, are associated with subsequent decreases in investment growth. Positive forecast errors following downward revisions are associated with increases in investment growth.

How should this evidence be interpreted? Managers may initially respond to analyst forecast revisions when making investment decisions. Then they adjust when realized earnings show that the forecasts were biased. This would be a true reversal driven by managers correcting initial mistakes. However it is also possible that both investment changes and forecast errors are responding to common underlying information. Again regularization creates predictable patterns in how analysts process that information. Our two stage procedure is an attempt to isolate the first channel by instrumenting forecast errors with forecast revisions. However, given the complexity involved in actual firm investment decisions, we cannot entirely rule out the second interpretation. Overall, we view Table \ref{tab:investment} as tentative evidence that the economic effects do matter for firm investments.


\section{Conclusion}
\label{sec:conclusion}

A large empirical literature documents systematic violations of rational expectations in professional earnings forecasts, macroeconomic predictions, and asset prices. The standard interpretation of this evidence attributes it to behavioral biases arising from cognitive limitations or emotional influences. We provide a rational, non-behavioral explanation with empirical work on professional analysts' earnings forecasts. Many of the patterns traditionally called underreaction and overreaction are predictable results of optimal statistical regularization.

To explain this, we developed a simple unified framework in which a forecaster minimizes mean squared prediction error in the presence of persistent fundamentals and transitory measurement noise. They optimally shrink the weight placed on new signals. This reflects proper recognition of the textbook bias–variance trade off. It generates forecast revisions and errors that systematically violate the \citet{coibion2015information} test, and also the levels test used by \citet{bordalo2020overreaction}, as well as related diagnostics. In our model the sign and magnitude of the apparent bias depend on (i) the intensity of regularization, and (ii) the relative persistence of fundamentals and noise. 

Our model provides monotonic comparative statics that map directly onto the heterogeneous patterns observed in the data. First, human analysts historically applied very strong implicit regularization, producing marked underreaction, leading to positive CG coefficients, especially in the short run when measurement noise is relatively small. This makes sense as there is a limit to how much time they have to evaluate a large number of minor effects. 

Second, purely statistical machine learning models such as ridge regression, gradient boosting, and random forests use more moderate, data determined regularization. At the short horizon, they largely eliminate overreaction. At a longer horizon, they generate massive overreaction with CG coefficients particularly low for ridge regressions. 

Third, direct experimental variation of the shrinkage parameter in simulations, confirms the central prediction. Stronger regularization monotonically increases the CG coefficient. That shifts forecasts from strong overreaction toward underreaction. 

Fourth, firms with noisier signals (younger, higher R\&D) have significantly more overreaction. This is exactly as implied by lower noise persistence and higher noise volatility. 

Fifth, the widespread adoption of accessible machine learning libraries around 2013 is associated with a sharp reduction in effective regularization as the machines reduce the cost of including many factors, and they help with finding optimal regularization. This is particularly marked among technically trained analysts. As a result, short-horizon underreaction disappeared and long-horizon overreaction surged. This is precisely the pattern predicted when regularization is somewhat reduced.

Jointly these results show that the range of logically possible combinations of who under-reacts or over-reacts more, are found in the data. This does not require standard behavioral biases. The evidence is a direct implication of rational forecasters (human or algorithmic) choosing different points on the bias–variance frontier; much as described in the textbooks \citep{hastie2009elements}.

Our model and our evidence has potentially broader implications. First, empirical rejections of forecast rationality using the standard tests should not be automatically interpreted as evidence of irrationality. What seems behavioral may simply be the statistical reflection of sensible and sophisticated forecasting methods. Second, as machine learning tools gain wider adoption, forecast biases will change in predictable ways as regularization becomes more transparent and directly controlled by ML algorithms. Third, because managers appear to take analyst forecasts seriously, the regularization induced prediction errors can have real effects on corporate investment.

Future research extending this framework to other forecasting environments, such as macroeconomic conditions and credit ratings, would be helpful. Developing diagnostic tools that distinguish rational regularization from genuine psychological bias is going to require more refined tests than are currently common. As we are in an era of increasingly algorithmic decision making, understanding the statistical origins of apparent behavioral anomalies may prove at least as important as documenting the anomalies themselves.

\bibliographystyle{chicago}
\bibliography{newdraft_refs}

\clearpage
\appendix

\setcounter{table}{0}
\renewcommand{\thetable}{A\arabic{table}}
\setcounter{figure}{0}
\renewcommand{\thefigure}{A\arabic{figure}}
\setcounter{equation}{0}
\renewcommand{\theequation}{A\arabic{equation}}
\setcounter{theorem}{0}
\renewcommand{\thetheorem}{A\arabic{theorem}}

\renewcommand{\thesection}{Appendix \Alph{section}}
\renewcommand{\thesubsection}{\thesection.\arabic{subsection}}

\section{Theoretical Derivations}
\label{appendix:model_proofs}

This appendix provides complete proofs of all propositions in the main text. We begin with preliminary results on AR(1) processes that are used throughout.

\subsection{Preliminaries: AR(1) Process Properties}

Consider a stationary AR(1) process.
\begin{equation}
x_t = \rho x_{t-1} + \varepsilon_t, \quad |\rho| < 1, \quad \varepsilon_t \sim \text{i.i.d.}(0, \sigma_\varepsilon^2).
\end{equation}

\begin{lemma}[AR(1) Moments]
\label{lem:AR1_moments}
For the stationary AR(1) process above
\begin{enumerate}
\item Unconditional variance: $\text{Var}(x_t) = \sigma_x^2 = \sigma_\varepsilon^2/(1-\rho^2)$
\item Autocovariance: $\text{Cov}(x_t, x_{t-k}) = \rho^k \sigma_x^2$ for $k \geq 0$
\item Variance of first difference: $\text{Var}(x_t - x_{t-1}) = 2(1-\rho)\sigma_x^2$
\item Covariance with first difference: $\text{Cov}(x_t, x_t - x_{t-1}) = (1-\rho)\sigma_x^2$
\end{enumerate}
\end{lemma}

\begin{proof}
(1) Taking variance of $x_t = \rho x_{t-1} + \varepsilon_t$ and using independence of $\varepsilon_t$ from $x_{t-1}$,
\begin{align}
\text{Var}(x_t) &= \rho^2 \text{Var}(x_{t-1}) + \text{Var}(\varepsilon_t) \\
\sigma_x^2 &= \rho^2 \sigma_x^2 + \sigma_\varepsilon^2 \\
\sigma_x^2 (1-\rho^2) &= \sigma_\varepsilon^2 \\
\sigma_x^2 &= \frac{\sigma_\varepsilon^2}{1-\rho^2}.
\end{align}

(2) We prove by induction. Base case $k=0$: $\text{Cov}(x_t, x_t) = \sigma_x^2 = \rho^0 \sigma_x^2$. Base case $k=1$.
\begin{align}
\text{Cov}(x_t, x_{t-1}) &= \text{Cov}(\rho x_{t-1} + \varepsilon_t, x_{t-1}) \\
&= \rho \text{Var}(x_{t-1}) + \text{Cov}(\varepsilon_t, x_{t-1}) \\
&= \rho \sigma_x^2 + 0 = \rho \sigma_x^2.
\end{align}
Inductive step: Assume $\text{Cov}(x_t, x_{t-k}) = \rho^k \sigma_x^2$. Then
\begin{align}
\text{Cov}(x_t, x_{t-k-1}) &= \text{Cov}(\rho x_{t-1} + \varepsilon_t, x_{t-k-1}) \\
&= \rho \text{Cov}(x_{t-1}, x_{t-k-1}) \\
&= \rho \cdot \rho^k \sigma_x^2 = \rho^{k+1} \sigma_x^2.
\end{align}

(3) Expanding
\begin{align}
\text{Var}(x_t - x_{t-1}) &= \text{Var}(x_t) + \text{Var}(x_{t-1}) - 2\text{Cov}(x_t, x_{t-1}) \\
&= \sigma_x^2 + \sigma_x^2 - 2\rho\sigma_x^2 \\
&= 2(1-\rho)\sigma_x^2.
\end{align}

(4) Expanding
\begin{align}
\text{Cov}(x_t, x_t - x_{t-1}) &= \text{Cov}(x_t, x_t) - \text{Cov}(x_t, x_{t-1}) \\
&= \sigma_x^2 - \rho\sigma_x^2 \\
&= (1-\rho)\sigma_x^2.
\end{align}
\end{proof}

\subsection{Ridge Coefficient}
\begin{lemma}[Ridge Coefficient for One Period]
The ridge coefficient with regularization parameter $\lambda$ at a one-period horizon is
\begin{equation}
\beta_{\lambda} = \frac{\alpha \rho_s \sigma_s^2}{\sigma_s^2 + \sigma_\eta^2 + \lambda}.
\end{equation}
\end{lemma}
\begin{proof}
The ridge estimator solves
\begin{equation}
\min_{\beta}  \mathbb{E}\big[(y_{t+1} - \beta z_{t})^2\big] + \lambda \beta^2,
\end{equation}
where $y_{t+1} = \alpha s_{t+1} + \epsilon_{t+1}$ and $z_{t} = s_{t} + \eta_{t}$. Differentiating the objective with respect to $\beta$ and setting equal to zero yields
\begin{equation}
\mathbb{E}[y_{t+1} z_{t}] = \beta \left(\mathbb{E}[z_{t}^2] + \lambda\right).
\end{equation}
Because $\mathbb{E}[y_{t+1} z_{t}] = \mathbb{E}[(\alpha s_{t+1} + \varepsilon_t) (s_{t} + \eta_{t)}]  = \alpha \rho_s \sigma_s^2$ and $\mathbb{E}[z_{t}^2] = \sigma_s^2 + \sigma_\eta^2$, the solution is
\begin{equation}
\beta_{\lambda} = \frac{\alpha \rho_s \sigma_s^2}{\sigma_s^2 + \sigma_\eta^2 + \lambda}.
\end{equation}
\end{proof}

\noindent The ridge coefficient for a two-period horizon can be calculated in similar ways.
\begin{lemma}[Ridge Coefficient for Two-Period Horizon]
The ridge coefficient at a two-period horizon is
\begin{equation}
\beta^{h=2}_{\lambda} = \frac{\alpha \rho_s^2 \sigma_s^2}{\sigma_s^2 + \sigma_\eta^2 + \lambda} = \rho_s \beta^{h=1}_{\lambda},
\end{equation}
where $\beta^{h=1}_{\lambda}=\frac{\alpha \rho_s \sigma_s^2}{\sigma_s^2 + \sigma_\eta^2 + \lambda}$.
\end{lemma}
\begin{proof}
The ridge estimator solves
\begin{equation}
\min_{\beta}  \mathbb{E}\big[(y_{t+1} - \beta z_{t-1})^2\big] + \lambda \beta^2,
\end{equation}
where $y_{t+1} = \alpha s_{t+1} + \varepsilon_{t+1}$ and $z_{t-1} = s_{t-1} + \eta_{t-1}$. Differentiating the objective with respect to $\beta$ and setting equal to zero yields
\begin{equation}
\mathbb{E}[y_{t+1} z_{t-1}] = \beta \left(\mathbb{E}[z_{t-1}^2] + \lambda\right).
\end{equation}
Because $\mathbb{E}[y_{t+1} z_{t-1}] = \mathbb{E}[(\alpha s_{t+1} + \epsilon_{t+1}) (s_{t-1} + \eta_{t-1})]  = \alpha \rho_s^2 \sigma_s^2$ and $\mathbb{E}[z_{t-1}^2] = \sigma_s^2 + \sigma_\eta^2$, the solution is
\begin{equation}
\beta^{h=2}_{\lambda} = \frac{\alpha \rho_s^2 \sigma_s^2}{\sigma_s^2 + \sigma_\eta^2 + \lambda}.
\end{equation}
\end{proof}

\subsection{Proof of Proposition \ref{prop:CG_optimal}}
\begin{proof}[Proof of Proposition \ref{prop:CG_optimal}]
Under optimal forecasting, $F_{t}y_{t+1} = \beta_\lambda z_{t}$ and $F_{t-1}y_{t+1} = \rho_s\beta_\lambda z_{t-1}$. The revision is
\begin{equation}
r_t = \beta_\lambda z_{t} - \rho_s\beta_\lambda z_{t-1} = \beta_\lambda(z_{t} - \rho_s z_{t-1}).
\end{equation}

Define $\tilde{\Delta}z_t = z_{t} - \rho_s z_{t-1}$. The CG coefficient is 
\begin{equation}
\gamma_{CG} = \frac{\text{Cov}(e_{t+1}, r_t)}{\text{Var}(r_t)} = \frac{\text{Cov}(e_{t+1}, \tilde{\Delta}z_t)}{\beta_\lambda \text{Var}(\tilde{\Delta}z_t)}.
\end{equation}

\textbf{Step 1: Computing $\text{Cov}(e_{t+1}, \tilde{\Delta}z_t)$.}

The forecast error is $e_{t+1} = y_{t+1}-F_{t-1}y_{t+1} = \alpha s_{t+1} - \beta_\lambda s_{t} - \beta_\lambda \eta_{t} + \epsilon_{t+1}$.

The persistence-adjusted difference is
\begin{align}
\tilde{\Delta}z_t &= z_{t} - \rho_s z_{t-1} \\
&= (s_{t} + \eta_{t}) - \rho_s(s_{t-1} + \eta_{t-1}) \\
&= (s_{t} - \rho_s s_{t-1}) + (\eta_{t} - \rho_s\eta_{t-1}).
\end{align}

Major observation. From $s_{t} = \rho_s s_{t-1} + v_{t}$, we have
\begin{equation}
s_{t} - \rho_s s_{t-1} = v_{t}.
\end{equation}
The fundamental component is exactly the innovation.

By linearity and independence,
\begin{align}
\text{Cov}(e_{t+1}, \tilde{\Delta}z_t) &= \alpha\text{Cov}(s_{t+1}, v_{t}) + \alpha\text{Cov}(s_{t+1}, \eta_{t} - \rho_s\eta_{t-1}) \\
&\quad - \beta_\lambda\text{Cov}(s_{t}, v_{t}) - \beta_\lambda\text{Cov}(s_{t}, \eta_{t} - \rho_s\eta_{t-1}) \\
&\quad - \beta_\lambda\text{Cov}(\eta_{t}, v_{t}) - \beta_\lambda\text{Cov}(\eta_{t}, \eta_{t} - \rho_s\eta_{t-1}) \\
&\quad + \text{Cov}(\epsilon_{t+1}, \tilde{\Delta}z_t).
\end{align}

By independence
\begin{equation}
\text{Cov}(e_{t+1}, \tilde{\Delta}z_t) = \alpha\text{Cov}(s_{t+1}, v_{t}) - \beta_\lambda\text{Cov}(s_{t}, v_{t}) - \beta_\lambda\text{Cov}(\eta_{t}, \eta_{t} - \rho_s\eta_{t-1}).
\end{equation}

\textbf{Step 2: Computing $\text{Cov}(s_{t+1}, v_{t})$.}

Since $s_{t+1} = \rho_s s_{t} + v_{t+1} = \rho_s(\rho_s s_{t-1} + v_{t}) + v_{t+1} = \rho_s^2 s_{t-1} + \rho_s v_{t} + v_{t+1}$
\begin{equation}
\text{Cov}(s_{t+1}, v_{t}) = \rho_s\text{Var}(v_{t}) = \rho_s\sigma_v^2 = \rho_s\sigma_s^2(1-\rho_s^2).
\end{equation}

\textbf{Step 3: Computing $\text{Cov}(s_{t}, v_{t})$.}

Since $s_{t} = \rho_s s_{t-1} + v_{t}$ and $v_{t}$ is independent of $s_{t-1}$:
\begin{equation}
\text{Cov}(s_{t}, v_{t}) = \text{Var}(v_{t}) = \sigma_v^2 = \sigma_s^2(1-\rho_s^2).
\end{equation}

\textbf{Step 4: Computing $\text{Cov}(\eta_{t}, \eta_{t} - \rho_s\eta_{t-1})$.}

\begin{align}
\text{Cov}(\eta_{t}, \eta_{t} - \rho_s\eta_{t-1}) &= \text{Var}(\eta_{t}) - \rho_s\text{Cov}(\eta_{t}, \eta_{t-1}) \\
&= \sigma_\eta^2 - \rho_s\rho_\eta\sigma_\eta^2 \\
&= (1-\rho_s\rho_\eta)\sigma_\eta^2.
\end{align}

\textbf{Step 5: Combining terms.}

\begin{align}
\text{Cov}(e_{t+1}, \tilde{\Delta}z_t) &= \alpha\rho_s\sigma_s^2(1-\rho_s^2) - \beta_\lambda\sigma_s^2(1-\rho_s^2) - \beta_\lambda(1-\rho_s\rho_\eta)\sigma_\eta^2 \\
&= (1-\rho_s^2)\sigma_s^2(\alpha\rho_s - \beta_\lambda) - \beta_\lambda(1-\rho_s\rho_\eta)\sigma_\eta^2.
\end{align}

\textbf{Step 6: Computing $\text{Var}(\tilde{\Delta}z_t)$.}

The fundamental component ($s_{t} - \rho_s s_{t-1}$) has variance
\begin{equation}
\text{Var}(s_{t} - \rho_s s_{t-1}) = \text{Var}(v_{t}) = \sigma_s^2(1-\rho_s^2).
\end{equation}

The noise component ($\eta_{t} - \rho_s\eta_{t-1}$) has variance
\begin{align}
\text{Var}(\eta_{t} - \rho_s\eta_{t-1}) &= \text{Var}(\eta_{t}) + \rho_s^2\text{Var}(\eta_{t-1}) - 2\rho_s\text{Cov}(\eta_{t}, \eta_{t-1}) \\
&= \sigma_\eta^2 + \rho_s^2\sigma_\eta^2 - 2\rho_s\rho_\eta\sigma_\eta^2 \\
&= (1 + \rho_s^2 - 2\rho_s\rho_\eta)\sigma_\eta^2.
\end{align}

By independence of fundamentals and noise
\begin{equation}
\text{Var}(\tilde{\Delta}z_t) = (1-\rho_s^2)\sigma_s^2 + (1+\rho_s^2-2\rho_s\rho_\eta)\sigma_\eta^2.
\end{equation}

\textbf{Step 7: Forming the ratio.}

\begin{align}
\gamma_{CG} & = \frac{(1-\rho_s^2)\sigma_s^2(\alpha\rho_s - \beta_\lambda) - \beta_\lambda(1-\rho_s\rho_\eta)\sigma_\eta^2}{\beta_\lambda[(1-\rho_s^2)\sigma_s^2 + (1+\rho_s^2-2\rho_s\rho_\eta)\sigma_\eta^2]} \\
& = \frac{(1-\rho_s^2)\sigma_s^2( \frac{\alpha\rho_s}{\beta_\lambda} - 1) - (1-\rho_s\rho_\eta)\sigma_\eta^2}{(1-\rho_s^2)\sigma_s^2 + (1+\rho_s^2-2\rho_s\rho_\eta)\sigma_\eta^2}. \\
\end{align}

Plug in
\begin{equation}
\beta_\lambda = \alpha \rho_s \frac{\sigma_s^2}{\sigma_s^2 + \sigma_\eta^2 + \lambda},
\end{equation}
we have
\begin{align}
\gamma_{CG}^{optimal}  & = \frac{(1-\rho_s^2)(\sigma_\eta^2 + \lambda) - (1-\rho_s\rho_\eta)\sigma_\eta^2}{(1-\rho_s^2)\sigma_s^2 + (1+\rho_s^2-2\rho_s\rho_\eta)\sigma_\eta^2} \\
& = \frac{(1-\rho_s^2) \lambda - \rho_s (\rho_s -\rho_\eta) \sigma_\eta^2}{(1-\rho_s^2)\sigma_s^2 + (1+\rho_s^2-2\rho_s\rho_\eta)\sigma_\eta^2}.
\end{align}

\end{proof}
\clearpage
\section{Variable Definitions}
\begin{table}[ht]
\centering
\caption{Variable Definitions}
\label{tab:variable_definitions}
\begin{threeparttable}
\small
\begin{tabular}{p{4.5cm}p{7.5cm}p{3cm}}
\toprule
Variable & Definition & Source \\
\midrule
\multicolumn{3}{l}{\textit{Panel A: Dependent Variables}} \\
EPS & Earnings Per Share, manually adjusted for the share split factor. & IBES Unadjusted Detail\\
Human Forecast & Consensus EPS Forecast, manually adjusted for share split factor & IBES Unadjusted Summary\\
ML Forecast & EPS forecast, calculated using predictors& \\
Forecast Error & Realized EPS minus Forecasted EPS & IBES Unadjusted Summary\\
One Year Forecast Revision & year t-1 Forecasted year t EPS minus year t-2 Forecasted year t EPS & \\
Two Year Forecast Revision & year t-2 Forecasted year t EPS minus year t-3 Forecasted year t EPS & \\

\midrule
\multicolumn{3}{l}{\textit{Panel B: Firm Characteristics}} \\
Investment Rate & capx/at & Compustat\\
Total Assets & at & Compustat\\
Age & Current Fiscal Year - First Fiscal Year Observed& Compustat\\
R\&D Expenditure & xrd/at& Compustat\\
Book-to-Market & Book/Market & WRDS Financial Ratios Suite\\
Return on Assets & Return on Assets & WRDS Financial Ratios Suite\\
Debt to Assets &Total Long Term Debt Plus Debt in Current Liabilities/Total Assets & WRDS Financial Ratios Suite\\
\midrule
\multicolumn{3}{l}{\textit{Panel C: Analyst Characteristics}} \\
Technical Background & Coded manually & \\
\bottomrule
\end{tabular}
\begin{tablenotes}
\small
\item This table provides detailed definitions and data sources for all variables used in 
the empirical analysis. All accounting variables are from Compustat. Analyst forecasts are 
from IBES. ML forecasts are constructed following \citet{zhang2025man}.
\end{tablenotes}
\end{threeparttable}
\end{table}
\clearpage

\begin{figure}[hbtp!]
	\caption{LinkedIn page}
	\label{fig:LinkedIn}
	\begin{center}
	\begin{minipage}{0.8\textwidth}
		\centering
		\includegraphics[width=1.0\linewidth]{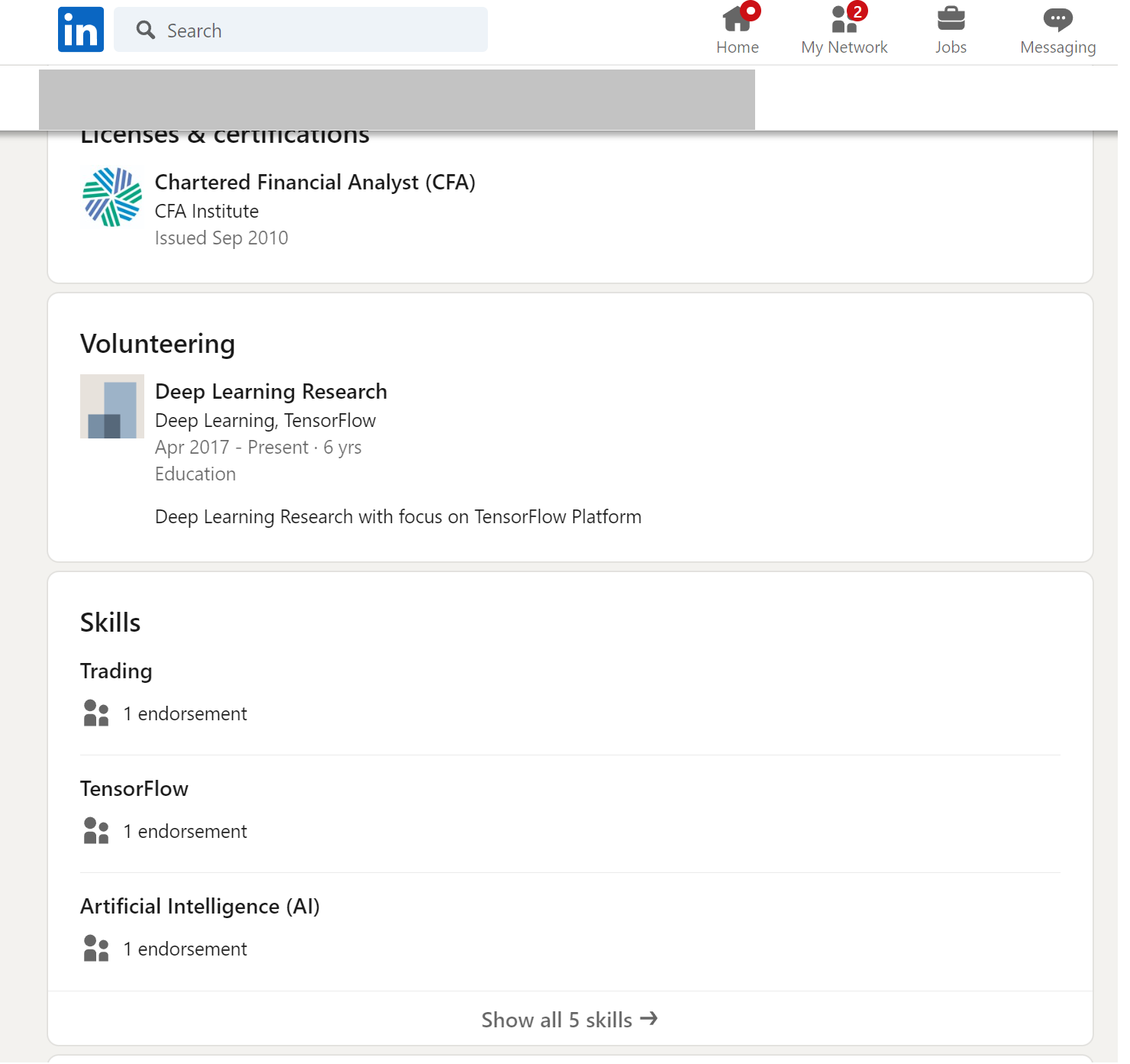}
	\end{minipage}
    \end{center}
    \begin{tablenotes}
    \centering
    \small
    \item This figure gives an example of a technical analyst's LinkedIn page.
    \end{tablenotes}
\end{figure}

\clearpage

\section{Calibration Details}
\subsection{Are These Effects Large Enough to Matter?}
To assess the quantitative relevance of our theoretical predictions, we calibrate the model to match key features of earnings forecast data. Table \ref{tab:calibration_results} reports predicted forecast efficiency test coefficients under both naive and optimal multi-period forecasting for economically relevant parameter configurations.

The calibration proceeds as follows. The noise-to-signal ratio $\sigma_\eta^2/\sigma_\varepsilon^2$ spans values from 0.5 to 2.0, corresponding to the interquartile range of annual earnings volatility (EPS scaled by lagged total assets) in Compustat over 1986--2019.\footnote{See \citet{dichev2009earnings} and \citet{dechow2002quality} for similar volatility ranges.} We set fundamental persistence $\rho_s = 0.9$ to match the average first-order autocorrelation of annual ROA for mature S\&P 500 firms with at least 20 years of data \citep[Table 3]{fama2006profitability}. For younger, high-growth firms in the bottom tercile of firm age, we use $\rho_s = 0.5$ \citep[Figure 1]{fama2006profitability}. Noise persistence $\rho_\eta = 0.5$ matches the average autocorrelation of IBES one-year-ahead forecast errors at the firm-year level in our sample (1994--2018), while $\rho_\eta = 0.9$ captures settings with highly persistent measurement error, such as recurring accounting restatements \citep[Table 4]{dechow2011predicting}. The regularization parameter $\lambda_{norm}$ represents the penalty normalized by total signal variance $(\sigma_s^2 + \sigma_\eta^2)$, ranging from 0 (no regularization) to 0.5 (strong shrinkage).

\subsection{Main Calibration Results}

Table \ref{tab:calibration_results} provides the main results. The calibration generates CG coefficients ranging from from $-0.327$ to $+0.069$. These ranges fully cover the empirical estimates in the literature. \citet{coibion2015information} report CG coefficients between $-0.10$ and $+0.20$ in macroeconomic forecasts. \citet{bordalo2020overreaction} reports values between $-0.15$ and $+0.30$ across different contexts. \citet{afrouzi2023overreaction} reports $\gamma_{CG} \approx -0.12$ for high-uncertainty firms. Our calibration manages to get this empirical coverage despite using only parameters directly estimated from earnings data. We do not need further free parameters chosen to fit forecast coefficients.

As predicted by the model, the CG coefficient depends critically and monotonically on the level of regularization. Across all specifications, an increase in regularization leads to an increase in the CG coefficient. In the specification with low noise volatility, the sign even shifts from negative (–0.216) to positive (0.069). When changing noise volatility—proxied by the noise-to-signal ratio scaled by the volatility of fundamentals—the CG coefficient also increases, consistent with the theoretical prediction. This result is also visualized by Figure 

Our calibration results show that, when the model is calibrated to match key features of earnings forecast data, it generates variation in the CG coefficient that is consistent in direction with the theoretical predictions, and the magnitude is economically meaningful.

\begin{table}[ht]
\centering
\caption{Calibration: Predicted Efficiency Coefficients}
\label{tab:calibration_results}
\begin{threeparttable}
\begin{tabular}{lcccc}
\toprule
 & \multicolumn{3}{c}{Parameters} &  \\
\cmidrule(lr){2-4} 
Scenario & $\rho_s$ & $\rho_\eta$ & $\lambda_{norm}$ & $\gamma_{CG}$ \\
\midrule
\multicolumn{4}{l}{\textit{Low Noise}} \\
 & 0.9 & 0.5 & 0.0 & -0.216   \\
 & 0.9 & 0.5 & 0.1 & -0.159   \\
 & 0.9 & 0.5 & 0.3 & -0.045  \\
 & 0.9 & 0.5 & 0.5 & 0.069  \\
\midrule
\multicolumn{4}{l}{\textit{Medium Noise}} \\
 & 0.9 & 0.5 & 0.0 & -0.279   \\
 & 0.9 & 0.5 & 0.1 & -0.235    \\
 & 0.9 & 0.5 & 0.3 & -0.147    \\
 & 0.9 & 0.5 & 0.5 & -0.058    \\
\midrule
\multicolumn{4}{l}{\textit{High Noise}} \\
 & 0.9 & 0.5 & 0.0 & -0.327    \\
 & 0.9 & 0.5 & 0.1 & -0.293    \\
 & 0.9 & 0.5 & 0.3 & -0.224    \\
 & 0.9 & 0.5 & 0.5 &  -0.155    \\
\bottomrule
\end{tabular}
\begin{tablenotes}
\small
\item This table reports predicted CG coefficients under both naive and optimal multi-period forecasting. $\lambda_{norm}$ is the regularization intensity normalized by total signal variance. $\gamma_{CG}$ is the Coibion-Gorodnichenko test coefficient. Parameters calibrated to match Compustat earnings volatility and IBES forecast patterns.
\end{tablenotes}
\end{threeparttable}
\end{table}
\clearpage

\clearpage

\begin{center}
	\Large \textbf{Internet Appendices for \textit{"Behavioral Machine Learning? Regularization and Forecast Bias"}} 
\end{center}

\vspace{1cm}
{\large
\begin{itemize}
	\item[] Internet Appendix A: Appendix Table
    \item[] Internet Appendix B: Machine Learning Prediction Detailed Steps
    \item[] Internet Appendix C: An Economic Model of Analyst Forecasts
    \item[] Internet Appendix D: Testing for Rational Expectations
\end{itemize}
}
\thispagestyle{empty}

\clearpage
\makeatletter

\setcounter{section}{0}
\renewcommand{\thesection}{Appendix \Alph{section}}
\renewcommand{\thesubsection}{\thesection.\arabic{subsection}}
\setcounter{table}{0}
\renewcommand{\thetable}{OA\arabic{table}}
\setcounter{figure}{0}
\renewcommand{\thefigure}{OA\arabic{figure}}
\setcounter{equation}{0}
\renewcommand{\theequation}{OA\arabic{equation}}
\setcounter{theorem}{0}
\renewcommand{\thetheorem}{OA\arabic{theorem}}


\appendixpagenumbering

\section{Appendix Table}\label{sec_app_table}
\begin{sidewaystable}[h]
\centering
\caption{Data Cleaning}
\begin{threeparttable}
\small
\begin{tabular}{llll}
\toprule
Step & Detail & Date & Obs\\
\midrule
\multicolumn{4}{l}{\textit{Construct Training Sample}} \\
1 & CRSP Data Monthly & 1980 Jan to 2023 Dec & 2,704,599\\
2 & Compustat Annual &  1964 to 2025 (fyear) & 496,758 \\
3 & Macro Data (RGDP RCON INDPROD UNEMP) &  1948 Jan to 2024 Apr  & 915 \\
4 & Financial Ratios Suite by WRDS &  1980 Jan to 2024 Dec (public\_date) & 2,619,622 \\
5 & IBES Actual EPS Detail Unadjusted Annual& 1976-01-15 to 2025-04-30 (ANNDATS) & 937,986 \\
6 & IBES EPS Summary Unadjusted Annual & 1980-01-31 to 2025-02-28 (FPEDATS) & 3,517,900\\
7 & Merge all file & 1984 Jan to 2019 Dec & 1,364,372 \\
\midrule
\multicolumn{4}{l}{\textit{Construct Final Sample}} \\
8 & Start with Merge all file & 1984 Jan to 2019 Dec (YearMonth) & 1,364,372 \\
9 & Conduct ML Training & 1984 Jan to 2019 Dec (YearMonth)& 1,312,517 \\
10 & Select Firm-Year Level Prediction & 1983 to 2020 (FPEDATS)& 124,491 \\
10 & Merge in Forecast Revision & 1984 to 2020 (FPEDATS) & 107,745 \\
11 & Keep Observation with NonMissing ML Forecast Error & 1985 to 2020 (FPEDATS) & 102,296 \\
12 & Keep Year from 1986 to 2019  & 1986 to 2019 (FPEDATS) & 101,737 \\
13 & Drop Firms with only 1 year observation  & 1986 to 2019 (FPEDATS) & 99,963 \\
\bottomrule
\end{tabular}
\begin{tablenotes}
\small
\item This table reports the data cleaning step.
\end{tablenotes}
\end{threeparttable}
\end{sidewaystable}

\clearpage


\begin{table}[h]
\centering
\caption{Machine Learning Predictors Definitions}
\label{tab:predictors}
\small
\begin{tabular}{p{2.5cm}p{5.5cm}p{2.5cm}p{5.5cm}}
\toprule
Variable & Definition & Variable & Definition \\
\midrule
\multicolumn{4}{l}{\textit{A. Firm Fundamental}} \\
\midrule
\multicolumn{4}{l}{\textit{IBES actual (1) }} \\
\cmidrule(lr){1-2} 
Last EPS & Last Year Actual EPS &   &   \\
\multicolumn{4}{l}{\textit{WRDS Financial Ratios (67) }} \\
\cmidrule(lr){1-2} 
accrual & Accruals\/Average Assets & intcov\_ratio & Interest Coverage Ratio \\
adv\_sale & Avertising Expenses\/Sales & inv\_turn & Inventory Turnover\\
aftret\_eq & After-tax Return on Average Common Equity & invt\_act & Inventor\/Current Assets\\
aftret\_equity & After-tax Return on Total Stockholders Equity & lt\_debt Long-term Debt\/Total iabilities\\
aftret\_invcapx & After-tax Return on Invested Capital & lt\_ppent & Total Liabilities\/Total Tangible Assets \\
at\_turn & Asset Turnover & npm & Net Profit Margin \\
bm & Book\/Market & ocf\_lct & Operating CF\/Current Liabilities \\
capei & Shillers Cyclically Adjusted P\/E Ratio  & opmad & Operating Profit Margin After Depreciation \\
capital\_ratio & Capitalization Ratio & opmbd & Operating Profit Margin Before Depreciation \\
cash\_conversion & Cash Conversion Cycle (Days) & pay\_turn & Payables Turnover \\
cash\_debt & Cash Flow\/Total Debt & pcf & Price/Cash fow \\
cash\_lt & Cash Balance\/Total Liabilities & pe\_exi & P\/E (Diluted, Excl. EI) \\
cash\_ratio & Cash Ratio & pe\_inc & P\/E (Diluted, Incl. EI) \\
cfm & Cash Flow Margin & PEG\_trailing & Trailing P\/E to Growth (PEG) ratio\\
curr\_debt & Current Liabilities\/Total Liabilities & pretret\_earnat & Pre-tax Return on Total Earning Assets \\
curr\_ratio & Current Ratio & pretret\_noa & Pre-tax return on Net Operating Assets \\
de\_ratio & Total Debt\/Equity & profit\_lct & Proft Before Depreciation\/Current Liabilities \\
debt\_assets & Total Debt\/Total Assets (1) & ps & Price/Sales\\
debt\_at & Total Debt\/Total Assets (2) & ptb & Price/Book\\
debt\_capital & Total Debt\/Capital & ptpm & Pre-tax Profit Margin\\
debt\_ebitda & Total Debt\/EBITDA & quick\_ratio & Quick Ratio (Acid Test)\\
debt\_invcap & Long-term Debt\/Invested Capital & RD\_SALE & Research and Development\/Sales\\
divyield & Dividend Yield & rect\_act & Receivables\/Current Assets\\
dltt\_be & Long-term Debt\/Book Equity & rect\_turn & Receivables Turnover\\
dpr & Dividend Payout Ratio & roa & Return on Assets\\
efftax & Effective Tax Rate & roce & Return on Capital Employed\\
equity\_invcap & Common Equity\/Invested Capital & roe & Return on Equity\\
evm & Enterprise Value Multiple & sale\_equity & Sales\/Stockholders Equity\\
fcf\_ocf & Free Cash Flow\/Operating Cash Flow & sale\_invcap & Sales\/Invested Capital\\
gpm & Gross Profit Margin & sale\_nwc & Sales\/Working Capital\\
GProf & Gross Profit\/Total Assets & short\_debt & Short-Term Debt\/Total Debt\\
\bottomrule
\end{tabular}
\end{table}

\begin{table}[h]
\ContinuedFloat
\centering
\caption{continued Variable Definitions}
\label{tab:predictors}
\small
\begin{tabular}{p{2.5cm}p{5.5cm}p{2.5cm}p{5.5cm}}
\toprule
Variable & Definition & Variable & Definition \\
\midrule
int\_debt & Interest\/ Average Long-term Debt & staff\_sale & Labor Expenses\/Sales\\
int\_totdebt & Interest\/Average Total Debt & totdebt\_invcap & Total Debt\/Invested Capital\\
intcov & After-tax Interest Coverage &&\\
\midrule
\multicolumn{4}{l}{\textit{CRSP (2) }} \\
\cmidrule(lr){1-2} 
ret & Monthly Return & prc & Closing Price \\
\midrule
\multicolumn{4}{l}{\textit{Fundamental Per Share (26, optional) }} \\
\cmidrule(lr){1-2} 
\multicolumn{4}{l}{\textit{A3. Firm Fundamental}} \\
\midrule
dividend\_p & Dividend Per Share & inventory\_p & Inventory Per Share\\
BE\_p & Book Equity Per Share & receivables\_p  & Receivables Per Share\\
Liability\_p & Total Liability Per Share & Cur\_debt\_p & Short-term Debt Per Share\\
cur\_liability\_p & Current Liabilities Per Share & interest\_p & Interest Per Share\\
LT\_debt\_p & Long-term Debt Per Share  & fcf\_ocf\_p & Free Cash Flow Per Share\\
cash\_p & Cash Balance Per Share & evm\_p & Enterprise Value Per Share\\
total\_asset\_p & Total Asset Per Share & sales\_p & Sales Per Share\\
tot\_debt\_p & Total Debt Per Share & invcap\_p & Invested Capital Per Share\\
accrual\_p & Accruals Per Share & c\_equity\_p & Current Equity Per Share\\
EBIT\_p & EBITDA Per Share & rd\_p  & RD Per Share\\
cur\_asset\_p & Current Asset Per Share & opmad\_p & Operating Income After Depreciation Per Share \\
pbda\_p & Operating Income before DA Per Share & gpm\_p & Gross Profit Per Share\\
ocf\_p & Operating Cash Flow Per Share & ptpm\_p & Pretax IncomePer Share\\
\midrule
\multicolumn{4}{l}{\textit{B. Macroeconomic variables (4)}} \\
RCON & Log Difference of Real Consumption  & INDPROD & Log Difference of Industrial Production Index\\
RGDP & Log Difference of Real GDP & UNEMP & Unemployment rate\\
\midrule
\multicolumn{4}{l}{\textit{C. Analyst forecasts (1)}} \\
AF & Analyst Forecasts&& \\
\bottomrule
\end{tabular}
\end{table}
\clearpage

\begin{table}[h]
\centering
\caption{Forecast Errors}
\label{onlinetab:errors}
\begin{threeparttable}
\small
\begin{tabular}{lllllllllll}
\toprule
\multicolumn{11}{l}{Panel A: Summary Statistics from Our Manuscript} \\
\midrule
   & RF    & AF    & AE    & RF-AE & t(RF-AE) & AF-AE & t(AF-AE) & (RF-AE)$^2$ & (AF-AE)$^2$ & Obs     \\
Y1 & 1.197 & 1.314 & 1.168 & 0.029 & 2.018    & 0.146 & 6.195    & 0.594    & 0.643    & 1246821 \\
Y2 & 1.368 & 1.731 & 1.367 & 0.001 & 0.016    & 0.364 & 8.28     & 1.79     & 1.839    & 1127459 \\
\midrule
\multicolumn{11}{l}{Panel A: Summary Statistics from Code Used in Zhang et al (2025)} \\
\midrule
   & RF    & AF    & AE    & RF-AE & t(RF-AE) & AF-AE & t(AF-AE) & (RF-AE)$^2$ & (AF-AE)$^2$ & Obs     \\
Y1 & 1.184 & 1.306 & 1.157 & 0.027  & 1.887  & 0.149 & 6.136 & 0.602 & 0.655 & 1265472 \\
Y2 & 1.356 & 1.727 & 1.359 & -0.004 & -0.072 & 0.368 & 8.261 & 1.805 & 1.856 & 1136233 \\
\bottomrule
\end{tabular}
\begin{tablenotes}
\small
\item This table tabulates the forecasting errors of Random Forests using a procedure closely aligned with the code published by \citet{zhang2025man}. Panel (a) reports our results, while Panel (b) reproduces the results without look-ahead bias from Panel C of Table 3, output cell 8, from \url{https://github.com/yandi-zhu/Man-versus-Machine-Learning-Revisited/blob/main/code/03_Main.ipynb}.
\end{tablenotes}
\end{threeparttable}
\end{table}

\clearpage

\begin{table}[h]
\centering
\caption{Machine Learning Forecast Errors Are Robustly Predictable}
\begin{threeparttable}
\small
\begin{tabular}{lcccc}
\toprule
\multicolumn{4}{l}{\textit{Panel A}} \\
& (1) & (2) & (3) & (4)  \\
& Human & Ridge & Gradient  & Random   \\
& Analysts & & Boosting  & Forest   \\
\cmidrule(lr){2-5}
& \multicolumn{4}{c}{$y_{i,t+1}   -F_{t} y_{i,t+1}$} \\
\midrule
$F_{t} y_{i,t+1}   -F_{t-1} y_{i,t+1}$ & 0.114*** & 0.006*** & 0.058*** & 0.044*** \\
                                       & (29.80)  & (2.71)   & (16.61)  & (12.84)  \\
\midrule
Firm FE & Yes & Yes & Yes & Yes  \\
Year FE & Yes & Yes & Yes & Yes  \\
Cluster at Firm & No & No & No & No \\
\midrule
N         & 99963    & 99963    & 99963    & 99963    \\
Adj R$^2$ & 0.16     & 0.11     & 0.11     & 0.10     \\
\midrule 
\multicolumn{4}{l}{\textit{Panel A}} \\
& (1) & (2) & (3) & (4)  \\
& Human & Ridge & Gradient  & Random   \\
& Analysts & & Boosting  & Forest   \\
\cmidrule(lr){2-5}
& \multicolumn{4}{c}{$y_{i,t+1}   -F_{t} y_{i,t+1}$} \\
\midrule
$F_{t} y_{i,t+1}   -F_{t-1} y_{i,t+1}$ & 0.122*** & 0.009*** & 0.064*** & 0.047*** \\
               & (15.13)  & (2.68)   & (7.46)   & (6.17)   \\
\midrule
Firm FE & Yes & Yes & Yes & Yes  \\
Year FE & No & No & No & No  \\
Cluster at Firm & No & No & No & No \\
\midrule
N              & 99963    & 99963    & 99963    & 99963    \\
Adj R$^2$      & 0.14     & 0.09     & 0.09     & 0.09     \\
\bottomrule
\end{tabular}
\begin{tablenotes}
\small
\item This table reports OLS regression results testing whether forecast errors are predictable using forecast revisions, following \citet{coibion2015information}. The dependent variable is the one-year-ahead forecast error, defined as realized minus one-year-ahead predicted EPS. The key independent variable is the change in forecasts from t-1 to t. Standard errors and fixed effects are indicated. *, **, and *** indicate significance at the 10\%, 5\%, and 1\% levels, respectively.
\end{tablenotes}
\end{threeparttable}
\end{table}
\clearpage

\begin{table}[h]
\centering
\caption{Machine Learning Forecast Errors Are Robustly Predictable at Two-Year Horizon}
\begin{threeparttable}
\small
\begin{tabular}{lcccc}
\toprule
\multicolumn{4}{l}{\textit{Panel A}} \\
& (1) & (2) & (3) & (4)  \\
& Human & Ridge & Gradient  & Random   \\
& Analysts & & Boosting  & Forest   \\
\cmidrule(lr){2-5}
& \multicolumn{4}{c}{$y_{i,t+2}   -F_{t} y_{i,t+2}$} \\
\midrule
$F_{t} y_{i,t+2}   -F_{t-1} y_{i,t+2}$ & -0.003    & -0.606*** & -0.233*** & -0.237*** \\
           & (-0.43)   & (-159.30) & (-37.30)  & (-38.90)  \\
\midrule
Firm FE & Yes & Yes & Yes & Yes  \\
Year FE & Yes & Yes & Yes & Yes  \\
Cluster at Firm & No & No & No & No  \\
\midrule
N         & 58864     & 58864     & 58864     & 58864     \\
Adj R$^2$ & 0.12      & 0.36      & 0.18      & 0.18      \\
\midrule
\multicolumn{4}{l}{\textit{Panel A}} \\
& (1) & (2) & (3) & (4)  \\
& Human & Ridge & Gradient  & Random   \\
& Analysts & & Boosting  & Forest   \\
\cmidrule(lr){2-5}
& \multicolumn{4}{c}{$y_{i,t+2}   -F_{t} y_{i,t+2}$} \\
\midrule
$F_{t} y_{i,t+2}   -F_{t-1} y_{i,t+2}$ & -0.004    & -0.496*** & -0.316*** & -0.319*** \\
           & (-0.31)   & (-20.58)  & (-28.58)  & (-29.40)  \\
\midrule
Firm FE & Yes & Yes & Yes & Yes  \\
Year FE & No & No & No & No  \\
Cluster at Firm & Yes & Yes & Yes & Yes  \\
\midrule
N          & 58864     & 58864     & 58864     & 58864     \\
Adj R$^2$  & 0.07      & 0.27      & 0.11      & 0.10      \\
\bottomrule
\end{tabular}
\begin{tablenotes}
\small
\item This table reports OLS regression results testing whether forecast errors are predictable using forecast revisions, following \citet{coibion2015information}. The dependent variable is the two-year-ahead forecast error, defined as realized minus two-year-ahead predicted EPS. The key independent variable is the change in forecasts from t-1 to t. Standard errors and fixed effects are indicated. *, **, and *** indicate significance at the 10\%, 5\%, and 1\% levels, respectively.
\end{tablenotes}
\end{threeparttable}
\end{table}

\clearpage

\begin{table}[h]
\centering
\caption{Subsample Analysis by Time Period }
\begin{threeparttable}
\small
\begin{tabular}{lcccc}
\toprule
\multicolumn{4}{l}{\textit{Panel A}} \\
& (1) & (2) & (3) & (4)  \\
& Human & Ridge & Gradient  & Random   \\
& Analysts & & Boosting  & Forest   \\
\cmidrule(lr){2-5}
& \multicolumn{4}{c}{$y_{i,t+2}   -F_{t} y_{i,t+2}$} \\
\midrule
$F_{t} y_{i,t+1}   -F_{t-1} y_{i,t+1}$ & 0.089*** & -0.011  & 0.028**  & 0.017    \\
                                       & (9.76)   & (-1.20) & (2.02)   & (1.35)   \\
\midrule
Firm FE & Yes & Yes & Yes & Yes  \\
Year FE & Yes & Yes & Yes & Yes  \\
Sample Period & 1986 - 2006 & 1986 - 2006 &1986 - 2006 & 1986 - 2006  \\
\midrule
N                                      & 63060    & 63060   & 63060    & 63060    \\
Adj R$^2$                              & 0.19     & 0.13    & 0.12     & 0.12     \\
\midrule
\multicolumn{4}{l}{\textit{Panel B}} \\
& (1) & (2) & (3) & (4)  \\
& Human & Ridge & Gradient  & Random   \\
& Analysts & & Boosting  & Forest   \\
\cmidrule(lr){2-5}
& \multicolumn{4}{c}{$y_{i,t+2}   -F_{t} y_{i,t+2}$} \\
\midrule
$F_{t} y_{i,t+1}   -F_{t-1} y_{i,t+1}$ & 0.091*** & 0.002   & 0.047*** & 0.030*** \\
                                       & (7.01)   & (0.68)  & (4.35)   & (3.00)   \\

\midrule
Firm FE & Yes & Yes & Yes & Yes  \\
Year FE & Yes & Yes & Yes & Yes  \\
Sample Period & 2007 - 2019 & 2007 - 2019 &2007 - 2019 & 2007 - 2019  \\
\midrule
N                                      & 36309    & 36309   & 36309    & 36309    \\
Adj R$^2$                              & 0.14     & 0.10    & 0.12     & 0.11     \\
\bottomrule
\end{tabular}
\begin{tablenotes}
\small
\item This table reports OLS regression results testing whether forecast errors are predictable using forecast revisions, following \citet{coibion2015information}. The dependent variable is the one-year-ahead forecast error, defined as realized minus one-year-ahead predicted EPS. The key independent variable is the change in forecasts from t-1 to t. Sampler periods are indicated. Standard errors (in parentheses) are clustered by firm. *, **, and *** indicate significance at the 10\%, 5\%, and 1\% levels, respectively.
\end{tablenotes}
\end{threeparttable}
\end{table}

\clearpage

\begin{table}[h]
\centering
\caption{Subsample Analysis by Time Period for Two-Year Horizon}
\begin{threeparttable}
\small
\begin{tabular}{lcccc}
\toprule
\multicolumn{4}{l}{\textit{Panel A}} \\
& (1) & (2) & (3) & (4)  \\
& Human & Ridge & Gradient  & Random   \\
& Analysts & & Boosting  & Forest   \\
\cmidrule(lr){2-5}
& \multicolumn{4}{c}{$y_{i,t+2}   -F_{t} y_{i,t+2}$} \\
\midrule
$F_{t} y_{i,t+2}   -F_{t-1} y_{i,t+2}$ & -0.045*** & -0.160*** & -0.189*** & -0.166*** \\
                                       & (-2.82)   & (-11.94)  & (-10.45)  & (-9.00)   \\
\midrule
Firm FE & Yes & Yes & Yes & Yes  \\
Year FE & Yes & Yes & Yes & Yes  \\
Sample Period & 1986 - 2006 & 1986 - 2006 &1986 - 2006 & 1986 - 2006  \\
\midrule
N                                      & 27771     & 27771     & 27771     & 27771     \\
Adj R$^2$                              & 0.15      & 0.14      & 0.15      & 0.13      \\
\midrule
\multicolumn{4}{l}{\textit{Panel B}} \\
& (1) & (2) & (3) & (4)  \\
& Human & Ridge & Gradient  & Random   \\
& Analysts & & Boosting  & Forest   \\
\cmidrule(lr){2-5}
& \multicolumn{4}{c}{$y_{i,t+2}   -F_{t} y_{i,t+2}$} \\
\midrule
$F_{t} y_{i,t+2}   -F_{t-1} y_{i,t+2}$ & -0.046**  & -0.690*** & -0.322*** & -0.328*** \\
                   & (-2.56)   & (-35.29)  & (-19.12)  & (-21.04)  \\

\midrule
Firm FE & Yes & Yes & Yes & Yes  \\
Year FE & Yes & Yes & Yes & Yes  \\
Sample Period & 2007 - 2019 & 2007 - 2019 &2007 - 2019 & 2007 - 2019  \\
\midrule
N                                      & 30513     & 30513     & 30513     & 30513     \\
Adj R$^2$                              & 0.15      & 0.48      & 0.24      & 0.25       \\
\bottomrule
\end{tabular}
\begin{tablenotes}
\small
\item This table reports OLS regression results testing whether forecast errors are predictable using forecast revisions, following \citet{coibion2015information}. The dependent variable is the two-year-ahead forecast error, defined as realized minus two-year-ahead predicted EPS. The key independent variable is the change in forecasts from t-1 to t. Sampler periods are indicated. Standard errors (in parentheses) are clustered by firm. *, **, and *** indicate significance at the 10\%, 5\%, and 1\% levels, respectively.
\end{tablenotes}
\end{threeparttable}
\end{table}

\clearpage

\begin{table}[ht]
\centering
\caption{Tech Analyst Coverage}
\label{tab:techcoverage}
\begin{threeparttable}
\small
    \begin{tabular}{lcccc}
    \toprule
    \multicolumn{5}{l}{\textit{Panel A: Coverage before and after 2013}} \\
     & (1)              & (2)              & (3)              & (4)              \\
     & Num of Stocks & Age       & Total   Assets & High Tech \\
    \midrule
    Tech*After 2013 & 0.722***  & -0.166    & -0.016    & -0.001    \\
            & (2.603)   & (-0.570)  & (-0.441)  & (-0.193)  \\
    \midrule
    Analyst FE      & Yes               & Yes       & Yes            & Yes         \\
    Year FE         & Yes               & Yes       & Yes            & Yes         \\
    Period          & 1994-2018         & 1994-2018 & 1994-2018      & 1994-2018   \\
    N               & 10132     & 10132     & 10132     & 10132     \\
    AdjR2          & 0.55      & 0.74      & 0.77      & 0.91      \\
    \midrule
    \multicolumn{5}{l}{\textit{Panel B: Tech Analysts Coverage}} \\
     & (1)     & (2)     & (3)    & (4)     \\
     & Num of Stocks & Age    & Total Assets & High Tech \\
    \midrule
    Tech        & -0.317*   & -1.788*** & -0.208*** & 0.226***  \\
            & (-1.768)  & (-7.156)  & (-6.288)  & (21.905)  \\
    \midrule
    Analyst FE & No        & No        & No        & No        \\
    Year FE    & Yes       & Yes       & Yes       & Yes       \\
    Period     & 1994-2018 & 1994-2018 & 1994-2018 & 1994-2018 \\
    N          & 10146     & 10146     & 10146     & 10146     \\
    AdjR2      & 0.07      & 0.06      & 0.05      & 0.05      \\
    \bottomrule
    \end{tabular}
\begin{tablenotes}
\small
\item This table estimates average characteristics of the stocks that tech analysts at analyst-year level. Tech is a indicator variables equalling to 1 if the analyst is tech analyst, 0 otherwise. After 2013 is an indicator variable equaling to 1 if the forecast is made after 2013. Number of Stocks is the total number of stocks covered by the analyst. Age is the average age of the stocks covered by the analyst. Total assets is the average total assets of stocks covered by the analyst. High Tech is the percentage of high-tech stocks covered by the analyst. High-Tech stocks are firms with three digit SIC code of 283, 357, 366, 367, 382, 384, or 737.
\end{tablenotes}
\end{threeparttable}
\end{table}

\clearpage

\section{Machine Learning Prediction Detailed Steps}\label{sec_online_ml}
Our machine learning prediction model strictly follows the existing literature on earnings forecasts. This ensures replicability and allows our results to be jointly evaluated alongside other research questions. Our data and sample are the same as \citet{van2023man} and \citet{zhang2025man}.

\subsection{Data and Sample}
Our data construction mirrors the approach in \citet{van2023man} and \citet{zhang2025man} and
relies on four underlying sources:
\begin{enumerate}
    \item CRSP monthly stock-level files,
    \item IBES Unadjusted Summary and Detail files for analysts' consensus forecasts
          and reported earnings,
    \item The WRDS \textit{Financial Ratios Suite} for firm fundamentals, and
    \item Real-time macroeconomic releases from the Federal Reserve Bank of
          Philadelphia.
\end{enumerate}

The sample consists of U.S.\ common equities (CRSP share codes 10 and 11) listed on
the NYSE, AMEX, or Nasdaq (exchange codes 1--3). The sample period runs from
January 1984 to December 2019. We analyze analysts'
one-, two- and three-year-ahead annual EPS forecasts (FPI $=1,2,3$).

Following \citet{van2023man} and \citet{zhang2025man}, the predictors used in our earnings-forecasting exercise fall into three broad categories: firm-level variables, macroeconomic indicators, and analysts' forecasts. Only the first firm-level variable listed below suffers from
look-ahead bias. As \citet{zhang2025man}, when we use firm-level variables "Most recently realized earnings", we ensure that the most recent realized earnings we used are not subject to look-ahead bias. The full list of predictors are provide below in Table \ref{tab:predictors}. This table is taken directly from the Internet Appendix (Table A.1) of \citet{zhang2025man}; we reproduce it here for completeness.

\subsection{Model Specification and Implementation}
Each month $t$, we estimate a random forest model to forecast firms' earnings per share (EPS) at different horizons $h$. 
\begin{equation}
\label{eq:mltraining}
    F_t (Y_{i,t+h}) = f_t^h(X_{i,t}|\Gamma)
\end{equation}

Let $Y_{i,t+h}$ denote firm $i$'s realized earnings per share (EPS) at time $t+h$. Let $f(\cdot)$ represent the machine learning function with hyperparameters $\Gamma$ applied to the feature set $X_{i,t}$. The forecasting horizons we consider include one year (A1), two years (A2), and three years (A3).

At each year-month $t$, we train the model in Equation~\ref{eq:mltraining} using a 12-month rolling window (from $t-12$ to $t-1$) when forecasting one-year-ahead earnings. We then apply the trained model $\tilde{f}_t^{h}(\cdot \mid \Gamma)$ to the feature vector $X_{i,t}$ to predict EPS at time $t+h$, $F_t (Y_{i,t+h})$. For two-year-ahead forecasts, we use a 24-month rolling window, and for three-year-ahead forecasts, we use a 48-month rolling window.

Our alternative specification is consistent with \citet{zhang2025man}. We report the detailed forecast comparison in Table \ref{onlinetab:errors}, and the results are very similar.

\subsection{Software Package}
We use scikit-learn version 1.3.0. For Ridge Regression, we import Ridge from sklearn.linear\_model; for Gradient Boosting, we import GradientBoostingRegressor from sklearn.ensemble; and for Random Forest, we import RandomForestRegressor from sklearn.ensemble.

\clearpage

\section{An Economic Model of Analyst Forecasts}
\label{sec:econ_model}
In this section, we build an economic model, based on which we generate simulated data to understand the behavioral biases documented in the real data. 

\subsection{Setup}
Based on \cite{begenau2018big}, we build up a repeated static model with one firm. In each period, the firm has a one-period investment project whose payoffs depend on the realizations of both the aggregate and idiosyncratic profitability. At the beginning of each period, the aggregate profitability realizes and is public to all agents in the economy. Then the firm chooses the number of shares to issue in order to maximize its net revenue. The firm uses the raised capital to finance its one-period investment opportunity with a stochastic payoff affected by both idiosyncratic and aggregate shocks. Then investors make portfolio choice decisions. Importantly, investors do not know the idiosyncratic profitability of the firm's investment project. They receive private signals about the profitability from two types of analysts randomly, i.e. rational and biased analysts. Observing the private signal and equity price in the market, each investor updates her belief about idiosyncratic profitability and forms posterior belief. Investors' decisions are made based on their posterior beliefs. At the end of each period, payoffs and utilities are realized. In the next period, new investors arrive and the same sequence repeats. Figure \ref{fig:timeline} illustrates the timing within each period $t$.

\begin{figure}[ht]
	\caption{Order of Events Within Period $t$} 
	\label{fig:timeline}
	\setlist[itemize]{nosep, leftmargin=*}
	\begin{tikzpicture}[
		node distance = 0mm and 0.02\linewidth,
		box/.style = {inner xsep=0pt, outer sep=0pt,
			text width=0.2\linewidth,
			align=left, font=\footnotesize}
		]
		\node (n1) [box]
		{   \begin{itemize}
				\item Nature
				\begin{itemize}
					\item $A_t$ picked and announced 
					\item $y_t$ picked, not yet announced
				\end{itemize}
			\end{itemize}
		};
		\node (n2) [box, below right=of n1.north east]
		{   \begin{itemize}
				\item Firm 
				\begin{itemize}
					\item $x_t$ equity issue
				\end{itemize}
			\end{itemize}
		};
		
		\node (n3) [box, below right=of n2.north east]
		{   \begin{itemize}
				\item Analysts 
				\item Type R analysts 
				\begin{itemize}
					\item Are $\lambda$ of these
					\item Send $\eta_{R,jt}\ (=y_t +\xi_{R,jt})$
				\end{itemize}
				\item Type B analysts 
				\begin{itemize}
					\item Are $1- \lambda$ of these
					\item Send $\eta_{B,jt}\ (=y_t+\xi_{B,jt})$
				\end{itemize}	
			\end{itemize}
		};
		\node (n4) [box, below right=of n3.north east]
		{   \begin{itemize}
				\item Investors
				\begin{itemize}
					\item Uniform draw assigns an analyst
					\item Get $\eta$ signal from analyst 
					\item Form beliefs
					\item $q_{j,t}$ equity demand
				\end{itemize}
    \item Equity market clears
			\end{itemize}
		};
		\node (n5) [box, below right=of n4.north east]
	{   \begin{itemize}
			\item Payoffs
			\begin{itemize}
				\item $y_{t}$ revealed 
				\item Payoffs determined
			\end{itemize}
		\end{itemize}
	};
	
		\draw[thick, -latex]    (n1.north west) -- (n5.north east);
		\draw (n1.north) -- + (0,3mm) node[above] {$t_{start}$};
		\draw (n2.north) -- + (0,3mm);
		\draw (n3.north) -- + (0,3mm);
		\draw (n4.north) -- + (0,3mm);
		\draw (n5.north) -- + (0,3mm) node[above] {$t_{end}$};
	\end{tikzpicture}
\end{figure}

\textbf{Firm} There is one firm that chooses the number of shares $x_t$ to issue in order to raise funds for investment. The capital raised from each share is invested in the firm's investment project whose payoff is $A_ty_t$. $A_t$ is the aggregate profitability, and $y_t$ is the idiosyncratic profitability. We introduce $A_t$ to conceptually allow for aggregate economic state. Another advantage is that including aggregate profitability increases the dimension of predictors we could use in simulations, which mimics the high dimensionality of predictors in the real data. Otherwise, $A_t$ functions exactly the same as $y_t$ in our current model.\footnote{This still holds even when we have more than one firms with different ex post idiosyncratic shocks. As we will see below, this is because CARA utility does have wealth effect and investors face no purchasing constraints.} Because we are interested in the analysts' forecasts on the \textit{firm-level} profitability, we assume that $A_t$ is public information and is realized at the beginning of each period. The logarithm of $A_t$ follows an AR(1) process 
\begin{align}
    \label{eq:ExoAgg}
    log A_t = \delta_AlogA_{t-1} + \epsilon_{At},
\end{align}
with $\epsilon_{At}\sim N(0,\rho_A^{-1}),\ \delta_A>0$.

The idiosyncratic productivity $y_t$ is not in the firm's information set, that is, the firm does not know the realized value of $y_t$ when it makes decisions. We assume that $y_t$ follows an AR(1) process which is conditional normal.
\begin{align}
    \label{eq:ExoIdio}
    & y_t = (1-\delta)\mu_y +  \delta y_{t-1} + \epsilon_{yt},
\end{align}
with $\epsilon_{yt}\sim N(0,\rho_y^{-1}),\ \delta>0,\ \mu_y>0$.

As will become clearer later, from the firm's perspective, $y_t$ matters only through equity price $p_t$. Thus, it is investors' belief about $y_t$, not the actual value of $y_t$, that matters. In this economy, the firm understands the equilibrium behaviors of investors. Conditional on all information realized in the previous period and current period aggregate profitability $A_t$, the firm owner maximizes its current period net revenue from the sale of the firm. That is, she solves the following problem
\begin{align}
	& \max_{x_t} E[x_t p_t - \frac{\phi}{2}(\Delta x_t)^2|\mathcal{I}_{t-1}, A_t], \label{firm}
\end{align}
where $ \frac{\phi}{2}(\Delta x_t)^2$ is equity issuance cost and $\Delta x_t =  x_t- x_{t-1}$. $\{\mathcal{I}_{t-1}, A_t\}$ is the information set of the firm owner when she makes equity issuance decision, therein, $\mathcal{I}_{t-1}$ represents all the information realized prior to time $t$. 

Given that the firm is the only supplier of stocks, its equity issuance affects the price of the stock $p_t$. Thus, the firm makes decisions knowing that its issuance affects the price of each share.

\textbf{Analysts} There are two types of analysts with a total measure of 1. $\lambda\in [0, 1]$ of them are rational investors who can make unbiased predictions (type R), $1-\lambda$ are analysts with biased expectations (type B). We could also assume that the former investors are biased but to a lesser extent, and all the conclusions would be the same.\footnote{We can also think of the `rational' investors as having bias $\epsilon$ which is infinitesimal.}

\textbf{Investors} There is a continuum of investors with a measure of 1. Each investor is endowed with wealth $W_t$ and decides how to allocate her wealth between riskless and risky assets, taking prices as given. 

\textbf{Investor information} The way that investors understand information is key to the model. 
When making decisions, investors do not know the firm's idiosyncratic profitability $y_t$, but they have a common prior that $y_t\sim N(\mu_{yt},\rho_y^{-1})$ with $\mu_{yt} = E[y_t|\mathcal{I}_{t-1}, A_t] = (1-\delta)\mu_y + \delta y_{t-1}$. 

Investors update their beliefs on $y_t$ based on private signals they receive from analysts and the stock price. Before making investment decisions, each investor randomly draws an analyst forecast from the pool of analysts. By the law of large numbers, $\lambda$ investors end up receiving the forecast of type R analysts, and $1-\lambda$ from type B analysts. We call the former investors type R (rational) investors, and the latter type B (biased) investors. Investors do not know the exact type of signal they receive.

Investor $j$ only receives one private signal from one analyst: 1) if the signal is provided by a type R analyst, the signal is $\eta_{R,jt} = y_t +\xi_{R,jt}$ with $\xi_{R,jt} \sim N(0, \rho_{\eta }^{-1})$; 2) if the signal is provided by type B analysts, it is $\eta_{B,jt} = y_t +\xi_{B,jt}$ with $\xi_{B,jt}\sim N(\theta\delta\epsilon_{t-1}, \rho_{\eta }^{-1})$. $\xi_{R, jt}$ ($\xi_{B, jt}$) is the difference between the true profitability and the signal value, that is, the total forecast error, and is uncorrelated across investors. The mean of the error term of type R analysts' forecasts is normalized to 0, and that of type B analysts is $\theta\delta\epsilon_{t-1}$, which captures the behavioral bias. 

Investors do not know about the potential biasedness of signals and therefore believe the private signal they receive is $\eta_{jt} = y_t + \xi_{jt}$, where $\xi_{jt}$ follows a normal distribution with mean 0
\[\xi_{jt}\sim N(0, \rho_{\eta }^{-1}).\]

Apart from the private signals, investors also learn about $y_t$ from the market price $p_t$, we assume and later on verify that observing this price is equivalent to observing a normally distributed signal $\eta_{pt} = y_t + e_t$ with $e_t\sim N(0,\rho_{pt}^{-1})$.

Then by Bayes' law, the perceived posterior distribution of $y_t$ is
\begin{align}
	& y_t|\eta_{jt},\eta_{pt} \sim N(\mu_{jt}, \rho_t^{-1}),
\end{align}
where $\rho_t^{-1}\equiv(\rho_y+\rho_{\eta }+\rho_{pt})^{-1}$ and $\mu_{jt}\equiv E[y_t|\mathcal{I}_{t-1},A_t,\eta_{jt},p_t]
= \rho_t^{-1}(\rho_y\mu_{yt} + \rho_\eta \eta_{jt} + \rho_{pt} \eta_{pt})$.
\textit{If type B investors know that the the signal they receive is biased, their updated beliefs on the mean of $y_t$,  $\mu_{jt}$, should be $\rho_t^{-1}(\rho_y\mu_{yt} + \rho_\eta (\eta_{jt}- \theta\beta\epsilon_{t-1}) + \rho_{pt} \eta_{pt})$.} However, as investors do not know about the potential biasedness of analyst signals, after a positive shock $\epsilon_{t-1}$, $1-\lambda$ investors who have received type B signals overestimate the firm's profitability.

\textbf{Investor portfolio choice}
After forming expectations about the firm's profitability, investors decide how to allocate their wealth between riskless and risky assets, taking prices as given.
\begin{align}
	& \max_{q_{jt}} U_{jt} = \alpha E_{jt}[W_{jt}|\mathcal{I}_{t-1}, A_t,\eta_{jt},p_t] - \frac{\alpha^2}{2}V_{jt}[W_{jt}|\mathcal{I}_{t-1}, A_t,\eta_{jt},p_t],\label{investor}\\
	& s.t.\ W_{jt} = r_t(W_t-q_{jt}p_t) + q_{jt}A_ty_t,
\end{align}
where $\{\eta_{jt},p_t, \mathcal{I}_{t-1}, A_t,\eta_{jt},p_t\}$ is investors' information set.

Plug the budget constraint into investor's utility, the maximization problem transforms into
\begin{align}
	\max_{q_{jt}} U_{jt} 
	& =  \alpha E_j\left[r_tW_t + q_{jt}(A_ty_t - p_tr_t)|\mathcal{I}_{t-1}, A_t,\eta_{jt},p_t \right]
	- \frac{\alpha^2A_t^2}{2}q_{jt}^2V_j(y_t|\mathcal{I}_{t-1}, A_t,\eta_{jt},p_t).
	\label{demand}
\end{align}

\textbf{Asset market clearing condition}
The equity market clears:
\begin{align}
\label{eq:market_clearing}
	 \int^1_0 q_{jt}dj =  x_t + \tilde x_t,
\end{align}
where $\tilde x_t$ is noisy supply and is distributed normally $N(0,\rho_x^{-1})$. The introduction of noisy supply prevents fully revealing $y_t$ with a continuum of traders.

\subsection{Equilibrium}\label{sec:eqm}
\begin{definition}
An equilibrium is a stock price $\{p_t\}$, firm equity issuance $ x_t$, and investor demand for equity $\{q_{jt}\}_j$, such that:
\begin{enumerate}
    \item Exogeneous states evolve according to Equations \ref{eq:ExoAgg} and \ref{eq:ExoIdio}.
    \item Each period $t$, given all the information realized before $t$, $\mathcal{I}_{t-1}$, and current aggregate profitability $A_t$, the firm owner chooses equity issuance amount $ x_t$ to maximize the firm's net revenue as in Equation \ref{firm}.
    \item Then, each investor $j\in[0,1]$ receives a private signal about the firm's idiosyncratic profitability $y_t$, which is randomly drawn from a continuum of analysts (meaning that investor $j$ does not know the type of signal she receives). In this economy, there are two types of analysts: $\lambda$ type R analysts and $1-\lambda$ type B analysts. The former produces unbiased signals $\eta_{R,jt}$, while the signal from the latter $\eta_{B,jt}$ suffers from overreaction and model misspecification error. Importantly, investor $j$ does not know the biasedness of $\eta_{B,jt}$, and her perceived signal is denoted as $\eta_{jt}$.
    \item Given all previous information and current aggregate profitability shock $\{\mathcal{I}_{t-1}, A_t\}$, a common prior about the distribution of $y_t$, a perceived private signal $\eta_{jt}$ about the firm's profitability $y_t$ and stock price $p_t$, investor $j$ updates her perceived distribution of $y_t$ according to Bayes' rule. Based on her updated posterior belief, she chooses the amount of equity to purchase, $q_{jt}$, to maximize her expected utility as in Equation \ref{investor}.
    \item Asset market clears following Equation \ref{eq:market_clearing}. 
\end{enumerate}
\end{definition}

To solve the model, we first solve for investors' equity demand as a function of the firm's equity issuance. We begin by solving for the investors' demand as a function of share price. Then we express their demand as a function of equity supply using the asset market clearing condition. Lastly, we plug investors' demand function into the firm's maximization problem and derive the equilibrium solution. The equilibrium value of equity issuance $ x_t$ is:
\begin{proposition}\label{prop:x}
	In equilibrium, the number of shares issued is
	\begin{align}\label{eq:eissuance}
		 x_t 
  =\frac{\frac{A_t}{\rho_t r_t}[ (\rho_y+\rho_{\eta }+\rho_{pt})\mu_{yt} + \rho_{\eta }(1-\lambda)\theta\delta\epsilon_{t-1} ] + \phi x_{t-1}}
		{2\frac{\alpha A_t^2}{\rho_t r_t} + \phi},
	\end{align}
	where $\rho_t =(\rho_y+\rho_{\eta }+\rho_{pt})$, $\rho_{pt} = \frac{\rho_\eta^2}{\alpha^2A_t^2}\rho_x$, $\mu_{yt} = (1-\delta)\mu_y + \delta y_{t-1}$.
\end{proposition}

Equation \ref{eq:eissuance} gives the total shares of equity the firm issues in equilibrium. The first term in the numerator captures the effect of expected $y_t$ which combines the information from the prior belief, private signals and stock price. The remaining terms in the numerator represent the marginal equity issuance costs. The denominator captures the sensitivity of price to demand plus the marginal equity issuance cost. High risk aversion ($\alpha$) leads to a lower equilibrium issuance: The intuition is that from the demand side, when investors are more risk averse, the amount of risky assets they would like to hold is lower for a given price level. Besides, high equity issuance cost ($\phi$) leads to less equity issuance when previous equity issuance is low ($\phi x_{t-1}$ is small): The intuition is that from the supply side, when equity issuance is more costly, the firm is less willing to issue more equity for a given price level.

\subsection{Simulation Results}\label{sec:simu}
In this section, we conduct \citet{coibion2015information} test in data simulated based on the economic model above. The simulation results are presented in Table \ref{tab:regularization_strength}, where the baseline parameters used in the simulations are given in Table \ref{tab:param}.

\begin{table}[ht]
 \caption{Parameter Values for Simulation}
   This table shows the baseline parameter values used for the simulation results. The value of $\theta$ is chosen based on the diagnostic parameter used in the literature \citep{bordalo2025real}. The riskless rate $r$ is normalized to 1. In the baseline case, the fraction of rational analysts is set to $\lambda=1$, to illustrate that we could obtain a non-zero \citet{coibion2012can} test coefficient, $\gamma_{CG}$, even when all agents are rational. The rest of the parameters are not calibrated to match real-world data. Instead, they are chosen to test the mechanism of overreaction of ML forecasts. Specifically, parameters associated with `Per share return', $\mu_y$, $\phi$ and $\alpha$ are chosen to generate large variations in this economy and increase the sensitivity of the firm's investment to investors' expectations.
  \label{tab:param}  
        \begin{center}
\begin{tabular}{lcccccc}
        \toprule
       Per share return & $\delta_A$ & $\sigma(\epsilon_A), \rho_A^{-\frac{1}{2}}$ & $\delta$ & $\sigma(\epsilon_z), \rho_y^{-\frac{1}{2}}$ & $\sigma(\eta), \rho_\eta^{-\frac{1}{2}}$ & $\sigma(\tilde x), \rho_x^{-\frac{1}{2}}$\\
       \cmidrule(lr){2-7}
       &  0.50 & 1.00 &  0.50 & 20.00 & 0.80 & 1.00 \\
        \midrule
    Investor decision & $\theta$ & $\lambda$ & $\alpha$ & r & $\mu_y$ & \\
       \cmidrule(lr){2-6}
       &  1.00 & 1.00 & 0.50 & 1.00 & 10.00 & \\
       \midrule
 Firm decision & $\phi$ & & & & & \\
       \cmidrule(lr){2-2}
       &  0.10 & &  & & & \\       
    \bottomrule
        \end{tabular}
        \end{center}
\end{table}

\begin{table}[h]
\centering
\caption{CG Coefficients in Simulated Data}
\label{tab:regularization_strength}
\begin{threeparttable}
\small
    \begin{tabular}{rrrr}
    \toprule
          &       & \multicolumn{1}{c}{(1)} & \multicolumn{1}{c}{(2)} \\
          &       & \multicolumn{1}{l}{Average $\gamma_{CG}$} & \multicolumn{1}{l}{Percentage of positive $\gamma_{CG}$} \\
    \midrule
    \multicolumn{1}{l}{Baseline} &       & -0.0056 & 51\% \\
    \midrule
    \multicolumn{1}{l}{Panel (a): $\alpha$} & \multicolumn{1}{l}{50}    & -0.0140 & 49\% \\
    & \multicolumn{1}{l}{500}   & -0.0108 & 50\% \\
    & \multicolumn{1}{l}{2000}   & -0.0002 & 54\% \\
          & \multicolumn{1}{l}{6000}   & 0.0268 & 57\% \\
    \midrule
    \multicolumn{1}{l}{Panel (b): Noise in $x$} & \multicolumn{1}{l}{$\sigma_\varepsilon = 0.5$} & -0.0075 & 50\% \\
    & \multicolumn{1}{l}{$\sigma_\varepsilon = 1.5$} & -0.0590 & 46\% \\
    & \multicolumn{1}{l}{$\sigma_\varepsilon = 2$} & -0.4779 & 30\% \\
    \midrule
    \multicolumn{1}{l}{Panel (c): $\lambda$} & \multicolumn{1}{l}{0.5}   & -0.0073 & 50\% \\
          & \multicolumn{1}{l}{0}     & -0.0074 & 50\% \\
    \bottomrule
    \end{tabular}%
\begin{tablenotes}
\small
\item This table presents the \citet{coibion2015information} test coefficients estimated using data simulated from the economic model. For each set of parameters, we simulate 100 panels, each of which consists of 1000 firms spanning 300 periods (with burn-in periods of 201). For each panel, we estimate the \citet{coibion2015information} test coefficient, $\gamma_{CG}$, from the equation below
\begin{equation}
e_{t+1} = \gamma_0 + \gamma_{CG} \, r_t + \omega_t,
\label{eq:CG_test_appx}
\end{equation}
where $e_{t+1} = y_{t+1} - F_{t} y_{t+1}$ is the forecast error and $r_t = F_{t} y_{t+1} - F_{t-1} y_{t+1}$ is the forecast revision. The forecasts are made using ridge. Predictors include the idiosyncratic profitability $y$ in the previous period, aggregate profitability $A$ and investment $x$.\\
Column 1 shows $\gamma_{CG}$ (times 1000 for display) averaged across the 100 panels. Column 2 shows the percentage of $\gamma_{CG}$ that are positive. In the baseline scenario, the regularization strength $\alpha$ is set to 1000, and all analysts are rational ($\lambda$=1). Panel (a) examines the results when we change the strength of regularization used by ridge. To examine our hypothesis that noise in predictors leads to more negative $\gamma_{CG}$, in panel (b) we assume econometricians can only observe $x$ with an additive exogenous noise $\varepsilon^x$. Specifically, we assume that the noise term $\varepsilon_t^x$ is independent of $x_t$ and follows an AR(1) process in logs. When making forecasts, instead of using $x_t$, the predictor used by econometricians is $x_t+\varepsilon^x_t-e^{\frac{1}{2}\sigma_\varepsilon^2}$, where $log(\varepsilon^x_t)=\rho_\varepsilon log(\varepsilon^x_{t-1})+\sigma_\varepsilon\epsilon^x_t$ and $\epsilon^x_t\sim N(0,1)$. The term $-e^{\frac{1}{2}\sigma_\varepsilon^2}$ is included so that the average value of this predictor remains unchanged.  $\rho_\varepsilon$ is set to 0.9 so that the noise term is persistent. Panel (c) tests the results when we change the percentage of rational analysts in the economy. When $\lambda=0.5$, half analysts are rational, and when $\lambda=0$, all analysts are overreacting. The remaining parameter values are given in Table \ref{tab:param}. 
\end{tablenotes}
\end{threeparttable}
\end{table}

\clearpage

\section{Testing for Rational Expectations}
\label{app:levels_test}

This appendix clarifies the terminology, historical development, and empirical use of the ``levels test'' for rational expectations, as referenced in the main text. We distinguish the traditional Mincer-Zarnowitz regression from the \citet{coibion2015information} revision-based test and trace their backgrounds.

\subsection{Terminology and Framework}

The benchmark null hypothesis that forecasts are unbiased and efficiently incorporate all information available at the time of the forecast is known as \textbf{Full Information Rational Expectations (FIRE)}. The term ``FIRE test'' is not standard and should be avoided. Instead, the literature has developed two distinct regression-based approaches to evaluating forecast rationality.
\begin{itemize}
    \item \textbf{Levels Test} (also known as the \citet{MincerZarnowitz1969} test or \textbf{MZ regression}).
    \[
    y_{t+h} = \alpha + \beta F_{t+h|t} + \varepsilon_{t+h}.
    \]
    Under rational expectations, forecasts are both unbiased and efficient, implying $\alpha = 0$, $\beta = 1$, and $\varepsilon_{t+h}$ orthogonal to the information set $\mathcal{I}_t$. Most empirical applications test only $\alpha = 0$ and $\beta = 1$, which can reject either because of bias or inefficiency.

    \item The \citet{coibion2015information} (CG) Test
    \[
    FE_{t+h|t} = \alpha + \rho \big(F_{t+h|t} - F_{t+h|t-1}\big) + \varepsilon_t,
    \]
    where $FE_{t+h|t} = y_{t+h} - F_{t+h|t}$ is the ex post forecast error. Under FIRE, forecast revisions contain no predictable component of future forecast errors, so $\rho = 0$.
\end{itemize}
The levels test evaluates unconditional forecast rationality, while the CG test isolates the response to \textit{new information} through forecast revisions.

\subsection{Academic Origin}

The levels test originates from early work on forecast evaluation in macroeconomics.
\begin{itemize}
    \item \citet{Muth1961} formally defined rational expectations as the conditional expectation of future realizations given available information.
    \[
    F_{t+1|t} = \mathbb{E}[y_{t+1} | \mathcal{I}_t],
    \]
    which implies $y_{t+1} = F_{t+1|t} + \varepsilon_{t+1}$ with $\varepsilon_{t+1} \perp \mathcal{I}_t$. Although Muth established the theoretical benchmark, he did not propose a direct regression test.

    \item \citet{MincerZarnowitz1969} introduced the first empirical regression framework for assessing forecast rationality. They estimated
    \[
    A_{t+1} = \alpha + \beta P_{t+1|t} + u_{t+1},
    \]
    where $A$ is the actual outcome and $P$ is the prediction, and tested the joint hypothesis
    \[
    H_0: \alpha = 0, \quad \beta = 1.
    \]
    This specification, sometimes called the \textbf{Mincer-Zarnowitz regression}, is the origin of what is commonly referred to as the \textbf{levels test}.
\end{itemize}
Subsequent research \citep{KeaneRunkle1990} refined the econometric interpretation of these regressions and highlighted that rejections may reflect inefficiency, measurement error, or heterogeneity rather than bias alone.

The levels test was quickly adapted to financial and accounting forecasts in the late 1970s and early 1980s. \citet{BrownRozeff1978} and \citet{FriedGivoly1982} applied the MZ regression to analyst earnings forecasts and found systematic departures from the rational expectations benchmark. These results motivated a substantial literature on forecast bias, underreaction, and behavioral distortions in financial expectations.

How does the levels test compare to the CG Test? 
\begin{table}[h]
\centering
\begin{tabular}{lcc}
\toprule
\textbf{Feature} & \textbf{Levels (MZ) Test} & \textbf{CG Test} \\
\midrule
Regression & $y = \alpha + \beta F + \varepsilon$ & $FE = \alpha + \rho \Delta F + \varepsilon$ \\
Null (FIRE) & $\alpha=0$, $\beta=1$ & $\rho=0$ \\
Detects & Bias, scale error, inefficiency & Over- or underreaction to new information \\
Power & Lower (confounded by level shifts) & Higher (targets information flow) \\
Interpretation & Overall rationality & Response to new information \\
\bottomrule
\end{tabular}
\caption{Comparison of Levels (MZ) and CG Tests of Forecast Rationality}
\label{tab:levels_cg}
\end{table}
The CG test is thought to provide a more targeted diagnostic of informational frictions because it examines how forecast revisions map into subsequent forecast errors. This focus on the processing of new information is thought to give the CG test higher power against behavioral or informational deviations such as over- or underreaction.

So the levels test originates with \citet{MincerZarnowitz1969} as they adopted the the theoretical foundation established by \citet{Muth1961}. It remains a useful benchmark that is used to  assess unconditional forecast rationality. In our analysis we report results from both frameworks. However, we emphasize the CG test since it is more widely used currently. 

\subsection{Orthogonality Tests of Rational Expectations}\label{app:orthogonality}

Next we examine the orthogonality test of rational expectations. This is the requirement that forecast errors $FE_{t+h|t} = y_{t+h} - F_{t+h|t}$ are uncorrelated with any variable $z_t$ in the forecaster’s information set $\mathcal{I}_t$ at time $t$. This test is more general than both the levels (MZ) and CG tests and forms the foundational empirical implication of Full Information Rational Expectations (FIRE).

We trace its academic origin, distinguish it from bias and revision-based tests, and clarify its relationship to the \citet{coibion2015information} framework and our regularization mechanism.

\subsection{Definition and Regression Form}

\begin{definition}[Orthogonality Condition]
Under FIRE,
\[
\mathbb{E}[FE_{t+h|t} \cdot z_t] = 0 \quad \forall z_t \in \mathcal{I}_t
\]
\end{definition}

Empirically, this is tested via the regression.
\begin{equation}
FE_{t+h|t} = \gamma' z_t + \varepsilon_{t+h}, \quad H_0: \gamma = 0
\label{eq:ortho}
\end{equation}
where $z_t$ is a vector of observable proxies for information available at $t$ (e.g., lagged earnings, past revisions, macroeconomic variables).

This is a rather general test of forecast rationality. Rejection implies either inefficiency, bias, or misspecified information.

\subsection{Academic Origin}

The orthogonality condition is a direct consequence of the law of iterated expectations. It was formalized in the earliest empirical tests of rational expectations.
\begin{itemize}
    \item \citet{Muth1961}. While not proposing a regression, Muth’s definition of rational expectations as $\mathbb{E}[y_{t+1} | \mathcal{I}_t]$ immediately implies that forecast errors are mean-zero and orthogonal to $\mathcal{I}_t$.
    \item \textbf{Muth (1960, unpublished technical report)} and \citet{nerlove1963}. Early simulation-based tests check whether errors are uncorrelated with lagged variables.
    \item \citet{MincerZarnowitz1969} In their seminal chapter, they explicitly test orthogonality alongside the levels regression.
    \begin{quote}
    ``A necessary condition for optimality is that the forecast error be uncorrelated with any information available at the time the forecast is made.''
    \end{quote}
    They regress errors on lagged actuals and predictions, finding significant coefficients. This is an early rejections of rationality.
\end{itemize}
Thus, \citet{MincerZarnowitz1969} are the origin of both the levels test and the orthogonality test in regression form.

\subsection{Evolution in Finance and Macroeconomics}

The orthogonality test became a workhorse in forecast evaluation.
\begin{itemize}
    \item \citet{fama1970efficient}.  In asset pricing, tests whether abnormal returns (prediction errors) are orthogonal to public information.
    \item \citet{BrownRozeff1978} and \citet{obrien1988}. Test whether analyst forecast errors are predictable using past forecast revisions, firm characteristics, or macroeconomic data.
    \item \textbf{Survey of Professional Forecasters (SPF)} literature. Regress inflation or GDP forecast errors on lagged inflation, unemployment, etc.
\end{itemize}

\subsection{Connection to Levels and CG Tests}

The three tests form a hierarchy.
\begin{table}[h]
\centering
\begin{tabular}{lccc}
\toprule
\textbf{Test} & \textbf{Regression} & \textbf{Condition Tested} & \textbf{Origin} \\
\midrule
Orthogonality & $FE = \gamma' z_t + \varepsilon$ & $\gamma = 0$ & Mincer-Zarnowitz (1969) \\
Levels (MZ) & $y = \alpha + \beta F + \varepsilon$ & $\alpha=0$, $\beta=1$ & Mincer-Zarnowitz (1969) \\
CG & $FE = \rho \Delta F + \varepsilon$ & $\rho = 0$ & Coibion-Gorodnichenko (2015) \\
\bottomrule
\end{tabular}
\caption{Hierarchy of Rational Expectations Tests}
\label{tab:test_hierarchy}
\end{table}

\begin{itemize}
    \item \textbf{Orthogonality} $\subset$ \textbf{Levels}. The levels test is a special case where $z_t = (1, F_{t+h|t})$.
    \item \textbf{CG} $\subset$ \textbf{Orthogonality}. The CG test uses $\Delta F_{t+h|t}$ as a \textit{sufficient statistic} for new information. It is a high-powered orthogonality test when revisions proxy for news.
\end{itemize}

The CG test is a \textbf{sharper} orthogonality test because $\Delta F$ is endogenous to the forecaster’s information flow.

\subsection{Connection to Regularization (This Paper)}

Our paper is a break from this tradition. Our theoretical framework generates orthogonality violations even under optimal forecasting.

\begin{proposition}[Regularization Induces Orthogonality Failure]
Let $\hat{F}_{t+h|t}(\lambda)$ be a ridge-regularized forecast. Then
\[
\text{Cov}(FE_{t+h|t}, \Delta \hat{F}_{t+h|t}) \neq 0
\]
because shrinkage distorts the mapping from news to revisions.
\end{proposition}

Empirically, we test
\[
FE_{it} = \gamma \cdot \text{Predictors}_{i,t-1} + \varepsilon_{it}
\]
and find $\gamma \neq 0$ post-ML adoption—consistent with optimal regularization, not behavioral failure.

We suggest that orthogonality tests reject not because forecasters are irrational, but because they rationally trade bias for variance.

The orthogonality test saying that forecast errors must be unpredictable using information in $\mathcal{I}_t$—is the core empirical implication of rational expectations. The levels test and CG test are specialized versions. Our paper shows that all three tests reject under optimal regularization, offering a unified non-behavioral explanation for decades of empirical puzzles.
\clearpage

\end{document}